\documentclass[aps,prf,superscriptaddress,showpacs,aps,longbibliography]{revtex4-2}

\pdfoutput=1
\usepackage[titletoc,toc,title]{appendix}
\usepackage{amsmath,amssymb}
\usepackage{graphicx}
\usepackage{xcolor}
\usepackage{rotating}
\usepackage{bm}
\usepackage{setspace}
\usepackage{hyperref}
\usepackage{float}
\usepackage{soul}
\usepackage[T1]{fontenc}
\usepackage{bbold}
\usepackage{color}
\usepackage{braket}
\usepackage{physics}
\usepackage{subcaption}
\usepackage{scrextend}

\graphicspath{{figures/}}

\hypersetup{
colorlinks=true,
urlcolor= blue,
citecolor=blue,
linkcolor= blue,
bookmarks=true,
}

\renewcommand{\vec}[1]{\bm{#1}}
\newcommand{\dH}{{\delta \mathcal{L}}}

\DeclareMathOperator{\sgn}{sgn}

\begin{document}

% \preprint{APS/123-QED}

\title{Topology of rotating stratified fluids with and without background shear flow}

\author{Ziyan Zhu}
\affiliation{Stanford Institute for Materials and Energy Sciences,
SLAC National Accelerator Laboratory, Menlo Park, CA 94025, USA}
\email{ziyanzhu@stanford.edu}
\author{Christopher Li} 
\affiliation{Department of Physics, Brown University, Providence, RI 02912-1843, USA}
\author{J. B. Marston}
\affiliation{Brown Theoretical Physics Center and Department of Physics, Brown University, Providence, RI 02912-1843, USA}

\begin{abstract}
     Poincar\'e inertio-gravity modes described by the shallow water equations in a rotating frame have non-trivial topology, providing a new perspective on the origin of equatorially trapped Kelvin and Yanai waves. We investigate the topology of rotating shallow water equations and continuously stratified primitive equations with and without background shear flow. Continuously stratified fluids support waves are analogous to the edge modes of weak three-dimensional topological insulators.  Background shear flow not only breaks the Hermiticity and homogeneity of the system but also leads to instabilities. By introducing a gauge-invariant winding number, we show that singularities in the phase of the Poincar\'e waves of the unforced shallow-water equations and primitive equations persist in the presence of both horizontal and vertical shear flows. Thus the bulk Poincar\'e bands have non-trivial topology and we expect and confirm the persistence of the equatorial waves in the presence of shear along the equator where the Coriolis parameter $f$ changes sign. 
\end{abstract}
\maketitle

\section{Introduction} 

Oceanic and atmospheric waves share fundamental physics with topological insulators and the quantum Hall effect, and topology plays an unexpected role in the movement of the atmosphere and oceans \cite{delplace2017}.  Topology guarantees the existence of unidirectional propagating equatorial waves on planets with atmospheres or oceans. In particular, there is a topological origin for two well-known equatorially trapped waves, the Kelvin and Yanai modes, caused by the breaking of time-reversal symmetry by planetary rotation.  Coastal Kelvin waves have also been demonstrated to have a topological origin~\cite{venaille2020}; thus Kelvin's 1879 discovery of such waves~\cite{Thomson:1880bv} likely marked the first time that edge modes of topological origin were uncovered in any context (though the topological nature remained hidden).  Recently reanalysis observations of Poincar\'e-gravity waves in the stratosphere have been used to demonstrate the non-trivial topology of these waves\cite{xu2023topological}. In light of these discoveries, it is important to consider the generalization of the shallow water equations to the more general problem of continuously stratified fluids. At the same time, it is also crucial to consider fluids driven by shear flows and damped by friction. Such extensions bring greater realism to models of actual fluids both on Earth~\cite{Vallis2017} and on other planets~\cite{Hammond:2019hw}.  The extension to background shear may also pave the way to the treatment of nonlinearities through the use of the mean-field quasilinear approximation~\cite{Malkus1956,Fried1960,Herring1963} that self-consistently treats the interaction of waves with mean flows. 

The existence of topological edge modes can be understood, via the principle of bulk-interface correspondence, to be predicted by the non-trivial topology of bulk modes.  Bulk-interface correspondence has been invoked for the quantum Hall effect and topological insulators~\cite{hasan2010topological,qi2011topological} as well as for a variety of classical wave systems, including nanophotonics~\cite{wang2009observation,plotnik2014observation,skirlo2014multimode,skirlo2015experimental}, accoustics~\cite{peano2015,yang2015,he2016}, mechanical systems~\cite{Nash2015,huber2016}, continuum fluids~\cite{silveirinha2015,delplace2017,shankar2017,green2020,venaille2020} and  plasmas~\cite{parker2020topological,Parker:2020bs,fu2021topological}.  The principle is clearest for Hermitian systems.  Driving and dissipation however lead to non-Hermitian dynamics \cite{kraus2012topological,lohse2016thouless,bandres2018lasers,zilberberg2018photonic,pedro2019topological}. By continuity, weak damping and driving may be expected to only change the waves slightly, but what happens as the forcing increases?  Efforts have been put into the topological classification of non-Hermitian systems~\cite{shen2018topological,gong2018topological,zhou2019periodic,borgina2020}.  Whether or not bulk-interface correspondence continues to hold remains a central problem. It has been argued that traditional bulk-interface correspondence breaks down in non-Hermitian systems~\cite{lee2016anomalous,xiong2018}.  Alternatives to the topological Chern number have been proposed~\cite{yao2018edge,kunst2018biorthogonal, gong2018topological, yin2018geometrical,helbig2020generalized}. Non-Hermitian bulk-interface correspondence has also been explored experimentally~\cite{xiao2020, ghatak2020observation}.   Here, we show that the phase singularity in the bulk wavefunctions persists in the presence of shear flow. The phase of the bulk Poincar\'e modes exhibits a vortex or anti-vortex at the origin in the wavevector space, with a change in the phase winding number across the equator. We show that equatorial Yanai and Kelvin waves persist in the background shear, consistent with the continued applicability of the principle of bulk-interface correspondence in the non-Hermitian realm.  

The paper is organized as follows. A brief introduction to topology in the context of fluid systems is presented in Section~ \ref{topology}.  It includes references to some pedagogical reviews.  We derive the shallow water equations in the presence of shear and compare numerical and perturbative methods to find the wave spectrum in Section~\ref{sec:swe}. The continuously stratified primitive equations with and without shear are analyzed in the f-plane approximation in Section~\ref{sec:primitive} and the Chern number for each band is found following the procedure introduced in Ref. \cite{delplace2017}, demonstrating a correspondence with weak 3D topological insulators.
In Section~\ref{sec:instability}, we show that the system is unstable with both horizontal and vertical shear. 
In Section~\ref{sec:winding} we numerically calculate the winding number to demonstrate the topology of the bulk. We first show that bulk-interface correspondence holds in the case of spatially varying Coriolis parameters. (The reader may wish to look at Ref. \cite{xu2023topological} which attempts to make the topological concepts discussed here accessible to climate scientists and geophysical fluid dynamicists.)  We then show our main result, which is that bulk-interface correspondence also appears to hold as background shear is turned on and the dynamics become non-Hermitian. Discussion and concluding remarks are made in Section~\ref{sec:conclusion}.  Some details of the calculations are relegated to Appendices.  
 
\section{Topological Invariants and Bulk-interface Correspondence}
\label{topology}

Topology is the branch of mathematics concerned with the qualitative shapes of objects that remain unchanged under continuous deformations.  The topological equivalence of a donut and a coffee mug (both have a single hole) is a commonly mentioned example, as is the fact that an M\"obius strip cannot be made orientable without tearing the paper and that it is impossible to comb the spines of a hedgehog (the Hairy Ball Theorem).

% Another example, more closely connected to the phenomena discussed below, is known as the Hedgehog or Hairy Ball Theorem that says that it is impossible to comb the spines of a hedgehog (because there will always be at least one tuft). 

% Topology is a powerful tool that turns complicated problems into simple ones, and fluids offer some compelling illustrations of topology at work.  

Topology finds noteworthy applications in fluids. Vortex rings for instance show persistence that is rooted in topology.  The persistence of vortex rings was striking enough for William Thomson (Lord Kelvin) to attempt to develop a theory of atoms based upon vortex rings in the hypothetical aether.  Kelvin’s circulation theorem states that the circulation (the line integral of the fluid velocity) around a closed loop that is advected with the fluid and thus deformed by the internal motion remains constant in the absence of viscosity and forcing. Tornadoes, hurricanes, Jupiter’s red spot, and even cutoff low-pressure regions in Earth’s atmosphere and vortex loops in the ocean are all examples of persistent vortices.   

Topology may also be applied to more abstract mathematical spaces.  In work recognized by the 2016 Nobel Prize in Physics, David Thouless and his collaborators demonstrated that the quantized conductance of the integer quantum Hall effect can be understood mathematically in terms of the topology of complex-valued wavefunctions that live on a compact Brillouin zone \cite{Thouless.1982}.  The electrical Hall conductance is proportional to an integer Chern number that characterizes the topology of the wave functions.  This quantization has a physical interpretation as electrical currents that propagate around the boundary of the semiconducting material in discrete modes, modes that owe their existence to the principle of bulk-interface correspondence.  The principle states that non-trivial topology away from a boundary implies the existence of unidirectional waves trapped along the boundary.  The quantum of resistance, $h/e^2$, can be measured so precisely that it has now been adopted as the international standard of resistance.   

The topology of linearized wave equations is frequently quantified in terms of the Chern number \cite{hasan2011}.  See Refs. \onlinecite{Faure.2019,Parker:2020hf,Horsley2022,xu2023topological} for some pedagogical reviews.  However, the Chern number has a number of drawbacks.  In contrast to systems on spatial lattices (where the Chern number was first applied), for continuous systems such as fluids the Chern number need not be integer-valued as it depends on how an integral over the Berry curvature is regularized at high wavevectors.  This ambiguity can sometimes be avoided by compactification \cite{delplace2017,venaille2020}.  Our viewpoint here is that this is more of a mathematical problem than a physical one because at small scales dissipation becomes strong providing a natural (albeit non-Hermitian) regularization at high wavenumbers.  Ultimately at the smallest scales, the fluid description passes over to Hamiltonian molecular dynamics.  It is unclear how to extend the Chern number to systems with dissipation, driving, or nonlinearities -- all properties of real fluids.

By contrast, these ambiguities do not arise for a winding number invariant. To demonstrate the concept of the winding number, it is necessary to define the gauge-invariant, complex-valued quantity $\Xi$ in the frequency-wavevector space:
\begin{eqnarray}\label{Xi}
\Xi(k_x, k_y) \equiv h^{*}(k_x, k_y)~ v(k_x, k_y),
\end{eqnarray}
where $h$ is the height and $v$ is the meridional velocity.
Normal wave modes, which are defined only up to an overall phase and magnitude, have their overall phases cancel out in Eq.~\eqref{Xi}, leaving only the relative phase difference between $h$ and $v$ and making $\Xi$ gauge-invariant.

\begin{figure*}[tbh]
\centering
\noindent\includegraphics[width=.9\textwidth]{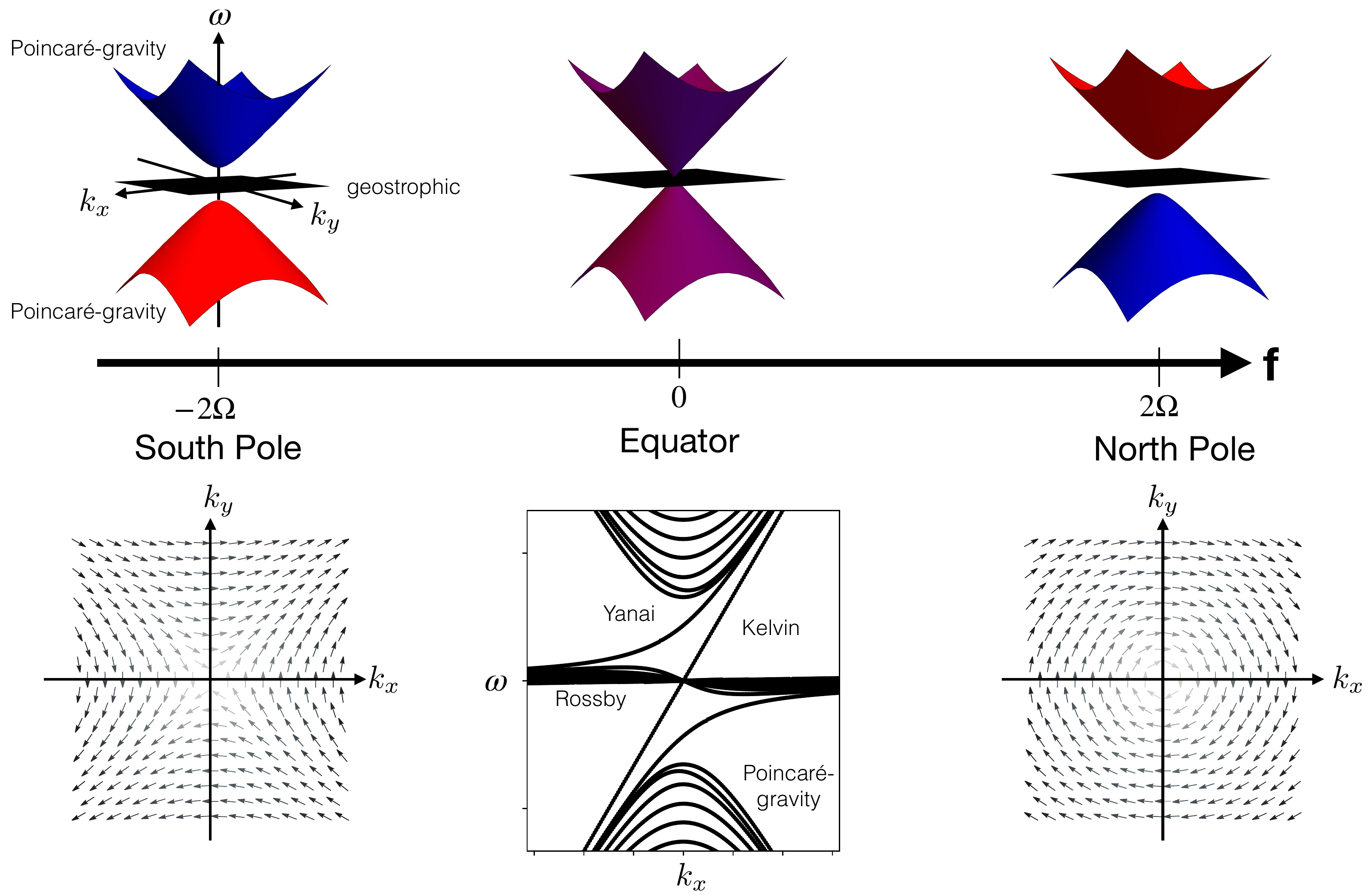}
\caption{
Dispersion relation in frequency-wavevector space of the rotating shallow water equations in the f-plane approximation as a function of latitude. The upper and lower bands are positive and negative frequency modes of the Poincar\'e waves, and the color indicates the sign of the winding number of the upper band (blue = -1, red = +1) as shown by the plots of the argument of $\Xi(k_x, k_y)$ in the lower half of the figure (see text). At the equator $f = 0$, the frequency gap vanishes in a Weyl point, and a topological transition occurs (purple) as the two bands invert.  The subinertial range has only a zero frequency band (black) containing modes in exact geostrophic balance. The inset shows the dispersion relation on the equatorial $\beta$-plane with the quasi-geostrophic Rossby waves, the Poincar\'e waves, and the unidirectional Kelvin and Yanai waves. The vectors plot corresponds to the south pole (left) and the north pole (right) respectively. (Figure and caption adapted from Ref. \cite{xu2023topological}.)}
\label{fig:windingnumber}
\end{figure*}

The idealized rotating shallow-water model on the f-plane is an illustration of the winding number. Figure~\ref{fig:windingnumber} shows the positive and negative frequency Poincar\'e modes (inertio-gravity waves) and the zero-frequency geostrophically balanced mode. The geostrophically balanced mode becomes Rossby waves if the Coriolis parameter varies with latitude. 
The topology of Poincaré-gravity modes is characterized by a vortex or antivortex in the frequency-wavevector space, with winding number $\pm 1$. A winding number of +1 means $\Xi$ increases (decreases) by $2\pi$ as one moves around the center of the vortex in a clockwise (counterclockwise) sense. On the other hand, the winding number of the geostrophic balanced mode is 0 (topologically trivial).
The winding number, as an alternative to the Chern number, serves the same function by quantifying the topology of the bands. 

A band inversion is a phenomenon where the winding number flips sign. This can occur for the Poincaré-gravity waves when either the Coriolis parameter or the wave frequency changes sign. According to the bulk-interface correspondence, the number of waves that traverse the otherwise forbidden region in the frequency space is the change in the winding number, which is 2 in this case.

Spectral flow in frequency-wavevector space as the zonal wavenumber increases shows that the negative frequency Poincar\'e band loses two modes, the geostrophic band gains and loses one mode and the positive frequency Poincar\'e band gains the two modes. These are the equatorial Kelvin and Yanai waves (The Yanai waves are also called mixed Rossby-gravity waves). The two equatorial modes move with an eastward group velocity at all zonal wavenumbers, and this unidirectional propagation is a consequence of the breaking of time-reversal invariance by the planetary rotation.

\section{Rotating shallow water equations with horizontal shear}~\label{sec:swe}
We begin this section by presenting the linearized rotating shallow water equations in the presence of horizontal shear and later consider vertical shear in the continuously stratified primitive equations.  (See Chapter 5 on ``Zonally symmetry wave -- mean interaction theory'' of Ref. \cite{buhler2014waves} for relevant background.)  
Note that $x$ and $y$ are zonal and meridional coordinates respectively. 
For simplicity we only consider shearing flow moving in the $x$-direction $\vec{U} (y) = (U(y), 0)$. 
We first introduce the following dimensionless quantities: 
\begin{eqnarray}
\tilde{t} = 2\Omega t, \quad \tilde{\eta} = \frac{\eta}{H},\quad \tilde{H}(y) = 1+\frac{h(y)}{H}, \quad \vec{\tilde u} = \frac{u}{c},\quad \vec{U} = \frac{U}{c}, \tilde{f} (y) = \frac{f(y)}{2\Omega},\quad  \vec{\tilde x} = \frac{\vec{x}}{L_d},
\end{eqnarray}
where $c = \sqrt{g H}$ is the gravity waves speed in nonrotating shallow water equations, $\Omega$ is the planet rotation rate, $H$ is zonally averaged depth in the absence of shear, and $L_d = c / 2\Omega$ is the global Rossby radius of deformation. 
Note that we assume $L_d$ is much smaller than the domain width, which allows us to treat the two equators independently. 
In terms of these nondimensionalized quantities and dropping the tildes for clarity, the shallow water equations after linearization and non-dimensionalization are given as follows (see Appendix~\ref{sec:derivation} for the derivation):
\begin{eqnarray}
&& \partial_{ t}  u + U(y) \partial_{ x}  u +  v \partial_{ y}  U(y) + \partial_{ x} {\eta} - f(y) v  = 0, \nonumber \\
&& \partial_{ t}  v + U(y) \partial_{ x}  v + \partial_{y}  \eta +  f(y)  u = 0,\nonumber \\
&& \partial_{{t}} {\eta} + {H}(y) (\partial_{{x}} {u} + \partial_{y}  v) +  v \partial_{ y}  H(y) + {U} (y) \partial_{x} 
{\eta} = 0, 
\label{eqn:swe_shear_mt}
\end{eqnarray}
where $u$, $v$ are respectively the $x$ and $y$ components of fluid velocity in the horizontal directions, $f(y)$ is the Coriolis parameter, $H(y)$ is the mean layer depth and $\eta$ is the fluctuation in the depth about this mean; thus the total layer depth is given by $h = H(y) + \eta$. 
Note that $H$ here is a function of $y$ due to the balance with the horizontal shear flow (see Eq.~\eqref{eqn:balance}) below).

We now further specialize to the case of a background basic shear flow that oscillates sinusoidally in the $y$-direction: 
\begin{eqnarray}
U(y) = U_0 \sin( \frac{2\pi y}{\Lambda} ), \label{eqn:uy}
\end{eqnarray}
where $U_0$ is the magnitude of the shear flow measured in units of $c \equiv \sqrt{g h}$ and $\Lambda$ is the wavelength of the shear. Note that linear shear $U(y) \propto y$ is incompatible with the periodic boundary conditions that we adopt in the following to eliminate any boundaries from the bulk problem that would confuse the application of the bulk-interface correspondence principle, as the only boundaries that we consider here are those located where the Coriolis parameter vanishes.    
Geostrophically balancing the basic flow then determines the mean depth $H(y)$, which satisfies:
\begin{eqnarray}
\pdv{H(y)}{y} = -f(y) U(y). \label{eqn:balance}
\end{eqnarray} 
In the f-plane approximation $f(y) = f_0$ for a constant $f_0$ and the mean depth is: 
\begin{eqnarray}
H(y) = 1 + \frac{U_0 f_0 \Lambda}{2\pi} \cos(\frac{2\pi y}{\Lambda}). \label{eqn:hy}
\end{eqnarray}

\subsection{Waves on a planet with two equators}
\label{TwoEquators}
\begin{figure*}[ht!]
\centering
    \includegraphics[width=0.6\linewidth]{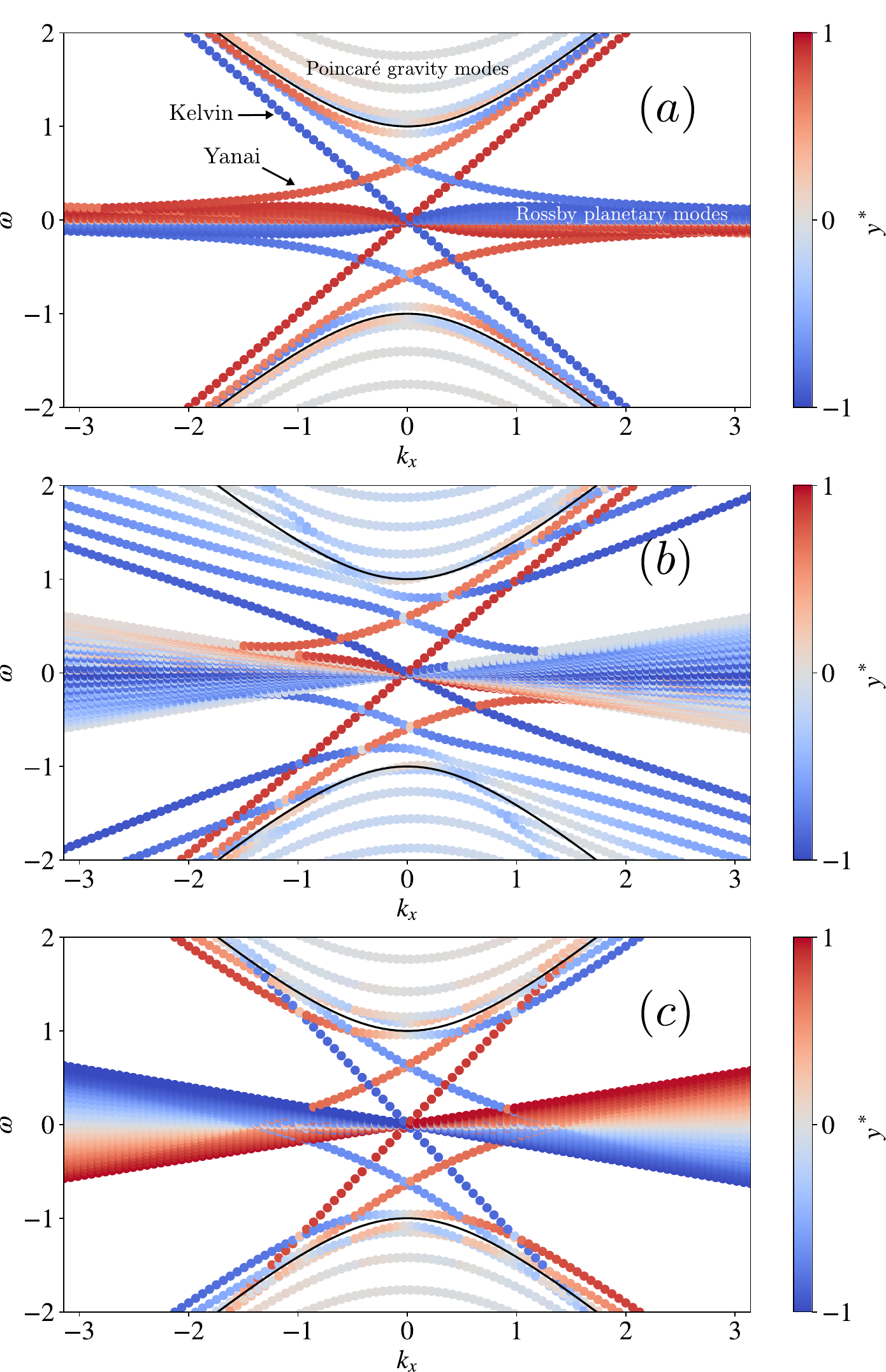} 
    \caption{Numerical evaluation of the frequency-wavenumber dispersion of the linearized shallow water equations obtained with \texttt{Dedalus} with $N_y = 61$ spectral modes showing the spectral flow of the Kelvin and Yanai waves between bands. 
    Colors show the projected real space position and $y^* = \langle \Psi | y | \Psi \rangle /L_y.$
    (a) No shear; (b) Imposed sine shear  (Eq.~\eqref{eqn:uy}) with $U_0 = 0.2$, and (c) Cosine shear (Eq.~\eqref{eqn:uy2}) with $U_0=0.2$. The Coriolis parameter varies sinusoidally (Eq.~\eqref{SinCoriolis}) and changes sign at $y=0$ ($y^*=-1$) and $y=\pm L_y/2$ ($y^*=1$). We set $L_y = 4\pi$. Black solid lines represent the frequency of the $k_y = 0$ Poincar\'e modes in the absence of shear and in the f-plane approximation $f=1$: $\omega = \pm \sqrt{k_x^2 + f^2}$. Colors represent the proximity of the band wavefunctions to the two equators. } 
    \label{fig:dedalus}
\end{figure*}

To investigate whether or not bulk-interface correspondence continues to hold in the presence of horizontal shear, we first examine 
the dispersion relation of shallow water waves in the presence of both rotation and shear. The wave frequencies are found numerically with the open-source pseudo-spectral \texttt{Dedalus} package~\cite{burns2019dedalus}. We employ $N_y = 61$ spectral modes in the $y$-direction, sufficient to resolve the waves and odd in number so that symmetry about $y = 0$ can be preserved.  We check that increasing the resolution $N_y$ does not change the frequencies significantly, including the Rossby wave frequency and the dispersion of the geostrophic modes.  We choose
\begin{eqnarray}
f(y) = \sin(\frac{2\pi y}{L_y})
\label{SinCoriolis}
\end{eqnarray}
and set $L_y = 4 \pi$ where $L_y$ is the width of the periodic domain (Fig.~\ref{fig:dedalus}). This choice respects the periodic boundary conditions and is sometimes called ``a planet with two equators'' as the Coriolis parameter changes sign twice across the domain~\cite{delplace2017}.   

Assuming the sinusoidal horizontal shear Eq.~\eqref{eqn:uy}, which is antisymmetric about the equator located at $y=0$, has the same periodicity as the domain size ($\Lambda = L_y$) the mean depth is: 
\begin{eqnarray}
    H(y) = 1 + U_0 \left[ \frac{L_y}{8\pi} \sin \left(\frac{4\pi y}{L_y} \right) -  \frac{y}{2} \right].
    \label{eqn:hy}
\end{eqnarray}
Similarly, if the shear is symmetric about the equator at $y = 0$, namely
\begin{eqnarray}
U(y) = U_0 \cos( \frac{2\pi y}{\Lambda} ), \label{eqn:uy2}
\end{eqnarray}
from geostrophic balance, the mean depth is:
\begin{eqnarray}
    H(y) = 1 + \frac{U_0 L_y}{8\pi} \cos \left(\frac{4\pi y}{L_y} \right)\ .
\end{eqnarray}
We consider both profiles in the following.  

In the absence of shear, Fig.~\ref{fig:dedalus}(a), 
equatorial Kelvin waves and Yanai waves appear in the gap between the high-frequency Poincar\'e and low-frequency planetary waves.  These waves have a topological origin~\cite{delplace2017}. 
They propagate unidirectionally (their group velocity does not change sign for all $k_x$), as guaranteed by topology.
Note that while the wave crest of the Rossby wave indeed always has a westward component, its group velocity can be both directions, as can be seen from the wave dispersion in Fig.~\ref{fig:dedalus}(a). 
As there are two oppositely-oriented equators, there are both eastward and westward propagating modes localized respectively at each equator. When shear $U_0 \neq 0$ is turned on, the planetary Rossby waves are Doppler shifted and we observe continuous spectra near the zero-frequency~\cite{brunet1990rossby}. The continuous spectrum spans $\omega = \pm U_0 k_x$. % to be continued 
The dispersion of the Poincar\'e modes also changes with increasing $k_x$; see Figs.~\ref{fig:dedalus}(b) and (c)). The Kelvin and Yanai waves remain localized near the equators. 
We have also investigated spectra with larger values of $U_0$ and find that the Kelvin and Yanai waves persist so long as $U_0$ is not too large. If $U_0$ is too large, the bulk bands and the boundary modes become difficult to distinguish due to significant changes in the frequency of the bulk modes and the large Doppler shift of the planetary waves, especially in the case of the sine shear flow. 
We show below that the continued presence of the waves is consistent with the persistence of bulk-interface correspondence in the presence of shear. 

\subsection{Bulk waves on the f-plane} \label{sec:analytic}
We now develop a purely spectral approach to including shear that is amenable to either direct diagonalization or a perturbative expansion.  First, we briefly review shallow water waves on the f-plane in the absence of shear flow~\cite{delplace2017}.
We expand the eigenmodes in the plane wave basis, $(u,v,\eta) = \Psi(k_x,k_y,f_0) = \hat{\Psi} \exp ( i k_x x + ik_y y - i \omega t)$.
In this basis, the linear wave operator is a $3 \times 3$ matrix:
\begin{eqnarray}
L_0 (k_x, k_y, f_0)  = \begin{pmatrix} 
 0 & i f_0 & k_x \\
 -i f_0 & 0 & k_y \\
 k_x & k_y & 0  \end{pmatrix}. \label{eqn:swe_h0}
\end{eqnarray}
% \begin{eqnarray}
% L_0 (k_x, k_y, f_0)  = \begin{pmatrix} 
%  0 & k_x &  k_y \\
%  k_x & 0 & -if_0 \\
%  k_y & if_0 & 0  \end{pmatrix}. \label{eqn:swe_h0}
% \end{eqnarray}
The amplitudes of the normal modes $\Psi_{\pm, 0}(k_x, k_y, f_0)$ with frequencies $\omega_{\pm, 0}$ can be obtained by diagonalizing $L_0$. The positive Poincar\'e mode frequency is $\omega_+ = \sqrt{k_x^2 + k_y^2 + f_0^2}$ with the eigenmode:
\begin{eqnarray} 
\Psi_+ = \begin{pmatrix} 
- \frac{k_x}{k} + i \frac{f_0 k_y}{k \sqrt{k^2 +f_0^2}}\\
\frac{k_y}{k} - i \frac{f_0 k_x}{k \sqrt{k^2+f_0^2}}\\
\frac{k}{\sqrt{k^2 + f_0^2}}
\end{pmatrix}, 
\end{eqnarray} 
where $k \equiv \sqrt{k_x^2 + k_y^2}$. A highly degenerate geostrophically-balanced mode appears at zero frequency, $\omega_0 = 0$ (the degeneracy is lifted when the Coriolis parameter varies with latitude or in the presence of shear):
\begin{eqnarray}
\Psi_0 (k_x, k_y, f_0) = \frac{1}{\sqrt{k^2 + f_0^2}} \begin{pmatrix} 
-ik_y\\ 
ik_x\\
f_0\end{pmatrix} 
\end{eqnarray} 
Finally the negative Poincar\'e mode has angular frequency $\omega_- = -\omega_+$ with corresponding wavefunction $\Psi_-(k_x, k_y, f_0) = \Psi_+(-k_x,-k_y,-f_0)$ reflecting the fact that the wave amplitudes in real space are real-valued. 

For Poincar\'e-gravity waves, the gauge-invariant quantity displays a vortex or antivortex (depending on the signs of the frequency and the Coriolis frequency) centered at the origin in wavevector space:  
\begin{equation}
 \Xi_{\pm}(k_x, k_y)= \frac{k_y-i \sgn(f_0) k_x }{ f_0},%i (k_x \mp i k_y \sgn(f_0)) / f_0\ .
\end{equation}
where we use the long-wavelength approximation $k^2 \ll f_0^2$. 
The vortex / antivortex has winding number $\pm 1$ which constitutes its topological charge.  Representing the phase of $\Xi$ with an arrow makes these patterns evident as shown in Figure \ref{fig:windingnumber}.   
The zero-frequency geostrophic mode, by contrast, has in the same limit 
\begin{equation}
 \Xi_{0}(k_x, k_y) = \frac{i k_x}{f_0},
 \label{Xi_0}
\end{equation}
and thus has a domain wall at $k_x = 0$ and zero winding number.  Its topological charge therefore vanishes.

\subsection{Horizontal Shear Flow on the f-plane}\label{sec:shear}
In the presence of shear flow the system is no longer translationally invariant along the $y$-direction. While the linear wave operator can still be expressed as a matrix in wavevector space, it is no longer composed of $3\times3$ block matrices along the diagonal. 
We first rewrite Eq.~\eqref{eqn:swe_shear_mt} in position space in the form of a matrix of differential operators, 
\begin{eqnarray}
\hat{L}(x, y, f_0, U_0) = i \begin{pmatrix} 
U(y) \partial_x & \pdv{U(y)}{y} - f_0 &  \partial_x \\ 
 f_0 &   U(y) \partial_x &  \partial_y \\ 
H(y) \partial_x &  H(y) \partial_y  - \pdv{H}{y} &  U(y) \partial_x
\end{pmatrix}.
\label{Lhat}
\end{eqnarray} 
This linear operator preserves the parity-time (PT) symmetry despite the broken Hermiticity, and spontaneous PT-symmetry breaking has been known to lead to instabilities~\cite{fu2020spontaneous,david2021cpt}. However, note that if the shear flow has dependence on both $x$ and $y$ (i.e., $U(x,y)$), PT symmetry would be broken. 
Substituting in the sine shear flow $U(y)$ from Eq.~\eqref{eqn:uy} with $H(y)$ satisfying the geostrophic balance in Eq.~\eqref{eqn:balance}, we obtain 
\begin{widetext}
	\begin{equation}
		\begin{split}
			\hat{L}(x, y, f_0, U_0) = i \begin{pmatrix} 
			U_0  \sin\left(\frac{2 \pi y}{\Lambda}\right)\partial_x & \frac{2\pi U_0}{\Lambda}  \cos\left(\frac{2\pi y}{\Lambda} \right) -  f_0 & \partial_x \\ 
			 f_0 &  U_0 \sin\left(\frac{2 \pi y}{\Lambda}\right) \partial_x  & \partial_y \\
			\left[ 1 + \frac{U_0 f_0 \Lambda }{2\pi}  \cos\left(\frac{2\pi y}{\Lambda} \right) \right] \partial_x & \left[1 + \frac{U_0 f_0 \Lambda}{2\pi} \cos\left(\frac{2\pi y}{\Lambda}\right) \right]  \partial_y - U_0 f_0\sin\left(\frac{2\pi y}{\Lambda}\right) & U_0 \sin\left(\frac{2 \pi y}{\Lambda}\right) \partial_x 
			\end{pmatrix}.\label{eqn:shear}
		\end{split}
	\end{equation} 
\end{widetext}
Note that the linear wave operator has a $y$-dependence, which means that when expanding $H$ in wavevector space, different modes with different $k_y$'s can mix. 
Without the loss of generality, we assume $\Lambda = 1$. We can consider the simplest case where there are only three modes, $k_y, k_y\pm 2\pi$, in the basis. In this case, the full linear wave operator is a $9 \times 9$ matrix that can be decomposed into $3\times 3$ blocks, which can be formally represented as follows, 
\begin{widetext}
\begin{equation}
\begin{split}
	\mathcal{L}_{9\times9}(k_x, k_y, f_0, U_0) = 
	\begin{pmatrix} 
		L_0(k_x, k_y + 2\pi, f_0) & T_1 (k_x, k_y, f_0, U_0) & 0 \\
		T_2 (k_x, k_y + 2\pi, f_0, U_0) & L_0 (k_x, k_y, f_0) & T_1(k_x, k_y - 2\pi, f_0, U_0) \\ 
		0 & T_2(k_x, k_y, f_0, U_0) & L_0(k_x, k_y-2\pi,f_0)
	\end{pmatrix}, \label{eqn:full}
\end{split}
\end{equation}
\end{widetext}
where $L_0$ is given in Eq.~\eqref{eqn:swe_h0} and $T_1$ and $T_2$ are the transition matrices between modes:
\begin{eqnarray} 
T_1(k_x, k_y, f_0, U_0) &=& \langle k_x, k_y + 2\pi | \hat{L} | k_x, k_y \rangle \nonumber \\
&=& \frac{U_0}{2} \begin{pmatrix}
i k_x & 2\pi i  & 0 \\ 
0 & i k_x & 0 \\
\frac{f_0 k_x}{2\pi} & \frac{k_y f}{2\pi} + f_0 & i k_x
\end{pmatrix}, \nonumber \\ 
T_2(k_x, k_y, f_0, U_0) &=& \langle k_x, k_y - 2\pi | \hat{L} | k_x, k_y \rangle \nonumber \\
&=& \frac{U_0}{2}  \begin{pmatrix}
-i k_x & 2\pi i  & 0 \\ 
0 & -i k_x & 0 \\
\frac{f_0 k_x}{2\pi} & \frac{k_y f_0}{2\pi} - f_0 & -i k_x 
\end{pmatrix}.\label{eqn:transition}
\end{eqnarray}
The derivation of $T_1$ and $T_2$ can be found in Appendix~\ref{sec:selection}. 
The matrix $T_1(k_x, k_y, f_0, U_0)$ connects wavenumber $k_y$ to $k_y + 2\pi$ and $T_2(k_x, k_y, f_0, U_0)$ connects $k_y$ to $k_y - 2\pi$. Note that $T_1 \neq T_2^\dagger$ and the linear wave operator is non-Hermitian. The frequency spectrum and the eigenvectors can then be obtained by diagonalizing the full matrix $\mathcal{L} (k_x, k_y, f_0, U_0)$. 
We validate our results by comparing our spectra with the ones obtained with \texttt{Dedalus}~\cite{burns2019dedalus} in Appendix~\ref{sec:dedalus}.

% The spectra for the cosine shear flow and an f-plane approximation can be obtained in a similar fashion, with $T_1$ and $T_2$ modified as follow: 
% \begin{eqnarray} 
% T_1(k_x, k_y, f, U_0) &=& \langle k_x, k_y + 2\pi | \hat{L} | k_x, k_y \rangle \nonumber \\
% &=& \frac{U_0}{2} \begin{pmatrix}
%  k_x & 2\pi   & 0 \\ 
% 0 &  k_x & 0 \\
% \frac{f k_x}{2\pi} & \frac{k_y f}{2\pi} + i f &  k_x
% \end{pmatrix}, \nonumber \\ 
% T_2(k_x, k_y, f, U_0) &=& \langle k_x, k_y - 2\pi | \hat{L} | k_x, k_y \rangle \nonumber \\
% &=& \frac{U_0}{2}  \begin{pmatrix}
%  k_x & -2\pi  & 0 \\ 
% 0 &  k_x & 0 \\
% -\frac{f k_x}{2\pi} & -\frac{k_y f}{2\pi} + i f & k_x 
% \end{pmatrix}.\label{eqn:transition}
% \end{eqnarray}
% The spectra are similar for sine and cosine shear with the f-plane approximation. 

\subsection{Perturbative treatment of shear} 
\begin{figure*}[ht!]
\centering
    \includegraphics[width=1.0\linewidth]{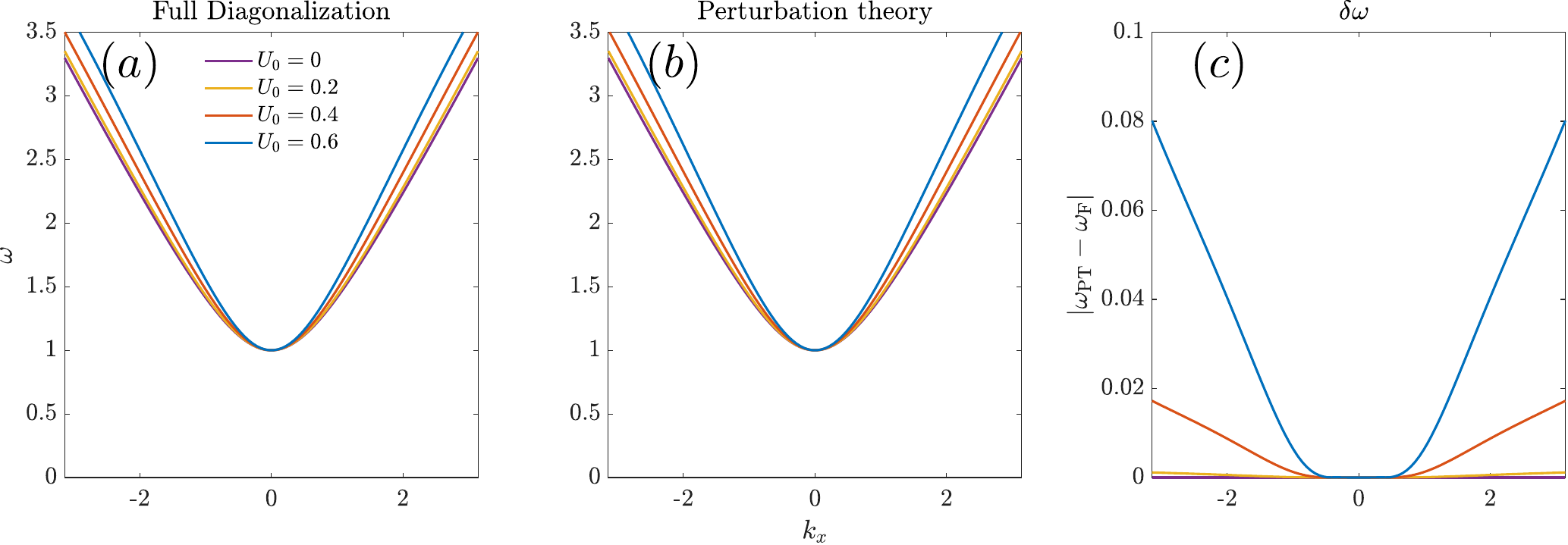} 
    \caption{Comparison of the frequency of the lowest positive frequency Poincar\'e modes from (a) full diagonalization and (b) perturbation theory of the $9\times9$ linear wave operator. (c) The difference between the two frequencies in (a) and (b).}
    \label{fig:pt}
\end{figure*}

We also consider a perturbative expansion of the eigenfunctions/values in powers of the shear~\cite{Sternheim1972, buth2004, boyd1978effects}.  
We may treat the shear flow perturbatively by considering the quantity $\delta \mathcal{L} = \mathcal{L} - \mathcal{L}_0$, namely the off-diagonal blocks in Eq.~\eqref{eqn:full}.
The correction to the frequency of the Poincar\'e mode first appears at second order in the shear: 
\begin{eqnarray}
\omega_n= \omega_n^{(0)} + \sum_{m\neq n} \frac{ \delta \mathcal{L}_{nm} \delta \mathcal{L}_{mn}}{\omega_n^{(0)} - \omega_m^{(0)}}, 
\end{eqnarray}
where $\delta \mathcal{L}_{mn} = \langle m | \delta \mathcal{L} | n \rangle$, and $m$ and $n$ are indices that label a wavevector state with some $k_y$. 
The wavefunctions including the first-order correction are given as follows: 
\begin{eqnarray}
| n \rangle = | n^{(0)} \rangle + \sum_{m\neq n} \frac{\dH_{mn} }{\omega^{(0)}_n - \omega^{(0)}_m} | m^{(0)} \rangle,
\end{eqnarray} 
where $|n^{(0)}\rangle$ and $|m^{(0)}\rangle$ are unperturbed wavefunctions corresponding to some $k_y$. 
The perturbed eigenmodes are still labelled by wavevector $(k_x, k_y)$ despite the fact that they contain contributions from modes at other $k_y$.  
To second order in the shear $U_0$, the frequencies only involve intermediate modes at wavevectors  $(k_x, k_y \pm 2\pi)$; higher orders of perturbations involve increasing departures of the wavenumber away from $k_x=0$.  
Figure~\ref{fig:pt} compares the frequency obtained from full diagonalization of the $9\times 9$ linear wave operator to the spectrum from second-order perturbation theory. The two spectra agree well with each other. 
Through comparing the perturbative spectrum and the full diagonalization, we show that firstly, the bulk can be classified by $(k_x, k_y)$ and secondly, the change of the bulk Poincar\'e wave is smooth as a function of $U_0$. 
Therefore, we argue that using the bulk-interface correspondence is valid despite the broken Hermiticity. 
As discussed later in Section \ref{sec:winding}, the first and second order perturbative corrections to the wavefunctions do not alter their topological properties.  

\subsection{Wave dynamics} 
\begin{figure}[ht!] 
\centering 
	\includegraphics[width=0.9\linewidth]{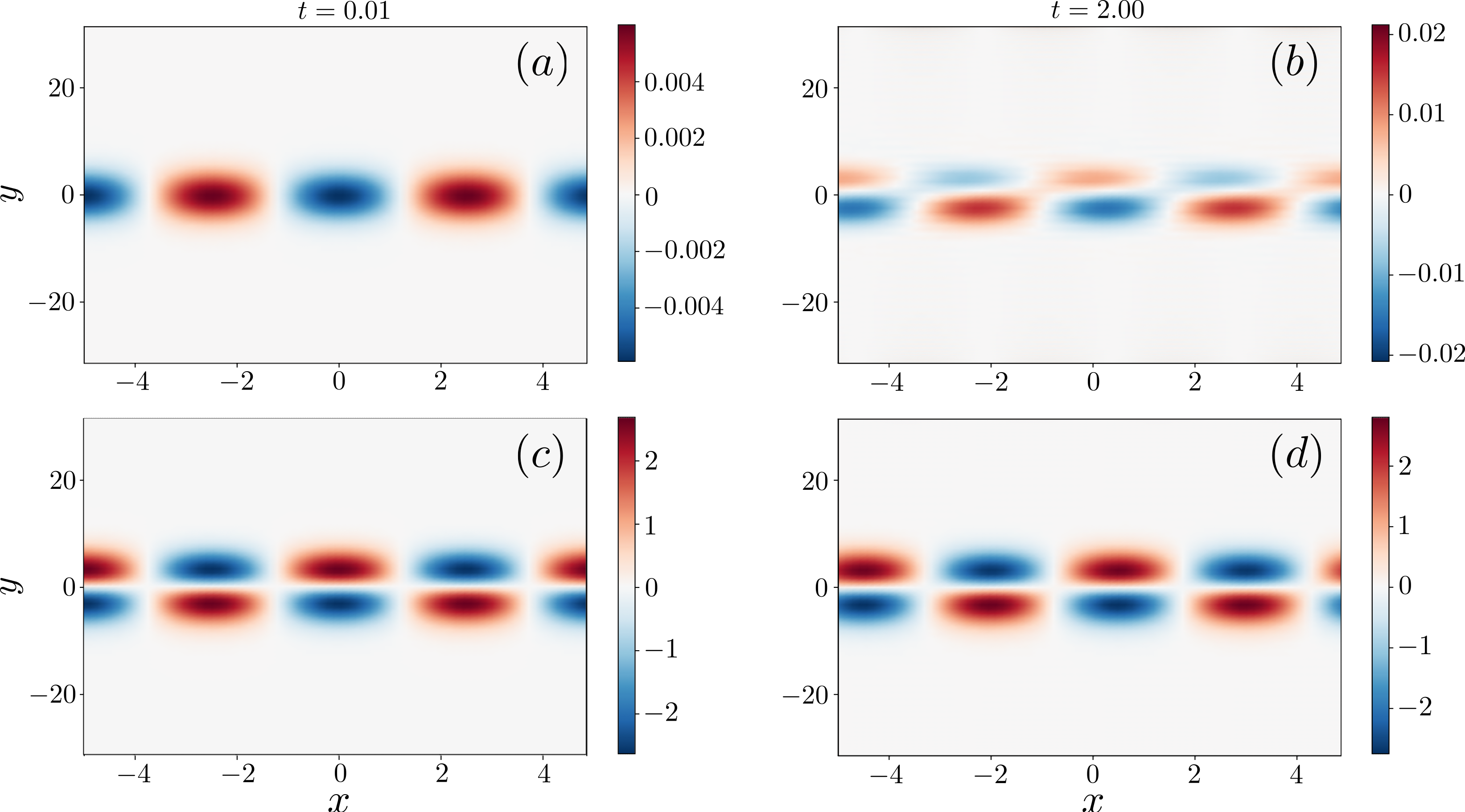}
	\caption{Time evolution of the $\eta$-component of (a, b) the Kelvin wave and (c, d) the Yanai wave for sine shear with $U_0=0.1$, $N_y =121$, $N_x=71$, $L_y=20\pi$, $L_x=10$. }
	\label{fig:sin_shear}
\end{figure}

\begin{figure}[ht!] 
\centering 
	\includegraphics[width=0.9\linewidth]{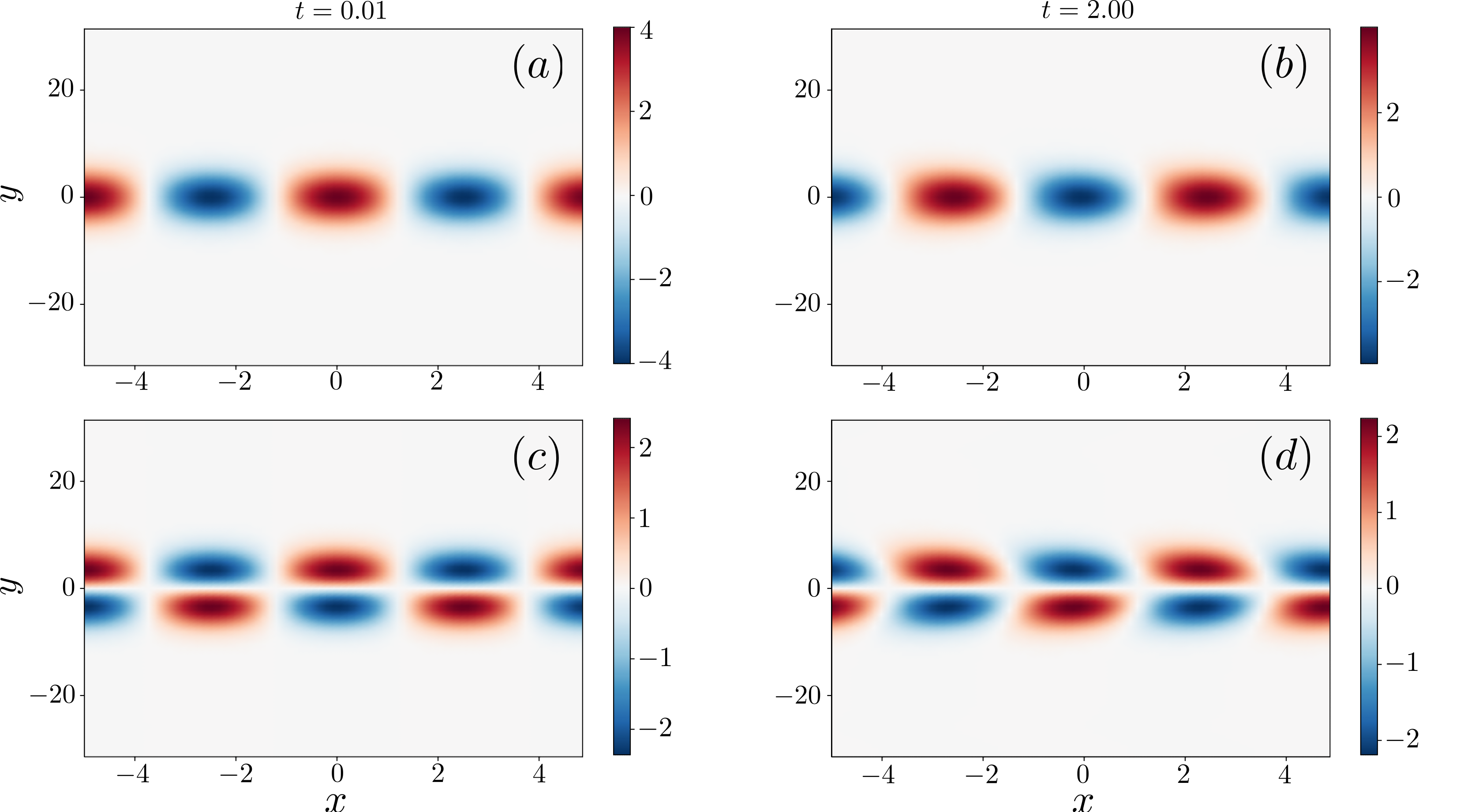}
	\caption{Time evolution of the $\eta$-component of (a, b) the Kelvin wave and (c, d) the Yanai wave for cosine shear with $U_0=0.1$, $N_y =121$, $N_x=71$, $L_y=20\pi$, and $L_x=10$. }
	\label{fig:cos_shear}
\end{figure}

Figures~\ref{fig:sin_shear} and \ref{fig:cos_shear} show snapshots of the propagation of wavenumber 2 ($k_x = 4 \pi / L_x$) Kelvin and Yanai waves subjected to sine and cosine shear. The waves remain localized near the $y = 0$ equator as they propagate. The wave amplitude grows in time with the sine shear (Figs.~\ref{fig:sin_shear} (b) and (d)) and decays in time with the cosine shear (Figs.~\ref{fig:cos_shear} (b) and (d)), consistent with the imaginary part of the frequency eigenvalues that correspond to growth and decay respectively for the two types of the shear. Note that since the sine shear is odd in $y$, the Kelvin wave also becomes asymmetric in $y$ as time evolves (Fig.~\ref{fig:sin_shear}(b)). 

\section{Primitive Equations With and Without Shear}\label{sec:primitive}
% without shear 
We turn next to the continuously stratified primitive equations. It has been shown that non-rotating stratified fluids with profiles of stratification that transition with increasing depth from marginally unstable to stable have a wave of topological origin along the interface~\cite{perrot2019transition}. % Our interest here, however, is to extend the topological theory of equatorial modes of the rotating shallow water equations to the case of purely stable and continuous vertical stratification. 
We make the standard Boussinesq approximation, and the vertical velocity or variation in the buoyancy replaces the depth as one of the dynamical fields. 

We first analyze the topological character of the linear stratified equations in the absence of shear by calculating the topological invariant within the bulk f-plane approximation. The linearized and non-dimensionalized equations can be derived from the underlying hydrostatic equations (see Appendix~\ref{sec:pe_w_shear} for the detailed derivation): 
\begin{eqnarray}
&& \frac{{\partial{u}}}{\partial{t}} = -U(y) \frac{\partial{u}}{\partial{x}} - v \frac{\partial{U(y)}}{\partial{y}} + f(y) v -  \frac{{\partial{\eta}}}{\partial{x}}, \nonumber \\
&& \frac{{\partial{v}}}{\partial{t}} =  -f(y) u - U(y) \frac{\partial{v}}{\partial{x}} -  \frac{{\partial{\eta}}}{\partial{y}},\nonumber \\
&& \frac{\partial}{\partial{t}} \frac{\partial{\eta}}{\partial{z}} = - w - U(y) \frac{\partial^2}{\partial{x} \partial{z}} \eta,
\label{eqn:primitive_eq}
\end{eqnarray}
where $w$ is the vertical velocity and the vertical depth variation $\eta$ and the buoyancy $b$ are related by the diagnostic relationship $\partial_z \eta = b$.

On the f-plane it is again natural to switch to a basis of plane waves which decouples the modes with different wavenumber in the z-direction, $k_z$. The incompressibility constraint in this basis takes the form
$\nabla \cdot \vec{u} = i(k_x u + k_y v + k_z w) = 0$ permitting the replacement of $w$ and $\eta$ with $b$, $u$ and $v$. In the absence of shear, Eq.~\eqref{eqn:primitive_eq} now corresponds in this basis to the linear wave operator
\begin{eqnarray} 
L_0  = \begin{pmatrix} 
 0 & i f_0 & -i \frac{k_x}{k_z} \\
 -i f_0 & 0 & -i \frac{k_y}{k_z} \\
  i\frac{k_x}{k_z} &  i \frac{k_y}{k_z} & 0  \end{pmatrix}\ . 
\label{eqn:stratified_matrix}
\end{eqnarray} 
% \begin{eqnarray} 
% L_0  = \begin{pmatrix} 
%  0 & i f_0 & k_x \\
%  -i f_0 & 0 & k_y \\
%   \frac{k_x}{k_z^2} &  \frac{k_y}{k_z^2} & 0  \end{pmatrix}. 
% \label{eqn:stratified_matrix}
% \end{eqnarray}
The eigenfrequencies of Eq.~\eqref{eqn:stratified_matrix} are $\omega_\pm = \pm \sqrt{f_0^2+ k^2/k_z^2}$ and $\omega_0 = 0$ with corresponding eigenvectors:
\begin{eqnarray}
\Psi_\pm &=& \frac{1}{\mathcal{N}_1}\begin{pmatrix} 
\mp i k_z k_x \sqrt{f_0^2 k_z^2 + k^2} + f_0 k_z^2 k_y\\
\mp i k_z k_y \sqrt{f_0^2 k_z^2 + k^2} - f_0 k_z^2 k_x \\
k^2 k_z
\end{pmatrix}, \nonumber \\ 
\Psi_0 &=&  \frac{1}{\mathcal{N}_2 }\begin{pmatrix} 
-k_x k_y \\ 
k_x^2 \\ 
f_0 k_x k_z 
\end{pmatrix},\label{eqn:primitive_dispersion}
\end{eqnarray}
where $k^2 = k_x^2 +k_y^2$, and $\mathcal{N}_{1,2}$ are normalization constants. 
%The Chern number can be calculated analytically by computing the integral of the Berry curvature over $k$ (odd viscosity can be used to regularize the integrand at large wavenumber).  
%We consider the positive frequency eigenvector at fixed non-zero $k_z$. The Berry connection is 

We consider the positive frequency eigenvector at fixed non-zero $k_z$. 
Letting $f_z = f_0 k_z$ and dividing $\Psi_{+}$ by $k_z$, we have
\begin{eqnarray}
\Psi_\pm &=& \frac{1}{\mathcal{N}_3}\begin{pmatrix} 
\mp i k_x \sqrt{f_z^2  + k^2} + f_z k_y\\
\mp i k_y \sqrt{f_z^2 + k^2} - f_z k_x \\
k^2
\end{pmatrix},
\end{eqnarray}
where $\mathcal{N}_3$ is the new normalization constant. The Berry connection is
\begin{eqnarray}
\mathrm{Im} \langle \Psi_+ | \grad_p | \Psi_+\rangle = (-2f_z k_y \sqrt{f_z^2+k^2}, 2f_z k_x \sqrt{f_z^2 +k^2},0),
\end{eqnarray}
where $\vec{p} = (k_x,k_y,f_z)$. The result is the same as the Berry connection of the positive Poincar\'e mode of the shallow water equations~\cite{delplace2017}. The difference of the Chern number between the two hemispheres, $\Delta C_{\pm}$, can be calculated analytically by integrating the Berry curvature over the unit sphere in $(k_x,k_y,f_z)$ space.  For the Poincar\'e modes, the difference $\Delta C_{\pm} = \pm 2$. 
By bulk-interface correspondence, for each $k_z$, there are two pairs of boundary Kelvin and Yanai modes (one pair each for the two oppositely oriented equators).  
%Note that there are bulk modes in between the positive and negative frequency Poincar\'e modes obtained from the f-plane approximation for $f=\pm 1$ (black solid lines in Fig.~\ref{fig:dedalus_pe}) that do not fill the entire gap. This is because of the slowing varying sinusoidal Coriolis parameter on the $L_y$ scale that takes values between -1 and 1. 
These stacks of boundary modes are analogous to the edge modes found in weak three-dimensional topological insulators~\cite{hasan2011}.

% \begin{figure*}[ht!]
% \centering
%     \includegraphics[width=1.0\linewidth]{{primitive_no_shear}.pdf} 
%     \caption{Spectral flow of Kelvin and Yanai waves between the band gaps exhibited by the linearized primitive equations obtained from \texttt{Dedalus} with $N_y = 61$ for the sinusoidal Coriolis parameter $f(y) = \sin(2\pi y/L_y)$ with $L_y = 5\pi$ and periodic boundary conditions in the horizontal directions.  The vertical wavenumber is (a) $k_z = 1$; (b) $k_z = 2$; and (c) $k_z = 3$. The solid black lines are the dispersion relation for the f-plane approximation with $f = 1$,  Eq.~\eqref{eqn:primitive_dispersion}. As in Fig.~\ref{fig:dedalus}, the color indicates proximity to the two equators. }
%     \label{fig:dedalus_pe}
% \end{figure*}

% with shear 
With sinusoidal horizontal shear flow, the eigenmodes of Eq.~\eqref{eqn:primitive_eq} can be obtained by the methods outlined in Section~\ref{sec:shear}. 
%Figure~\ref{fig:primitive spectrum} shows the eigenfrequency spectrum of the $69\times69$ matrix in comparison with the result obtained from \texttt{Dedalus}. 
The method again shows excellent agreement with the result obtained from \texttt{Dedalus} (not shown in the paper). 

\begin{figure*}[ht!]
\centering
    \includegraphics[width=1.0\linewidth]{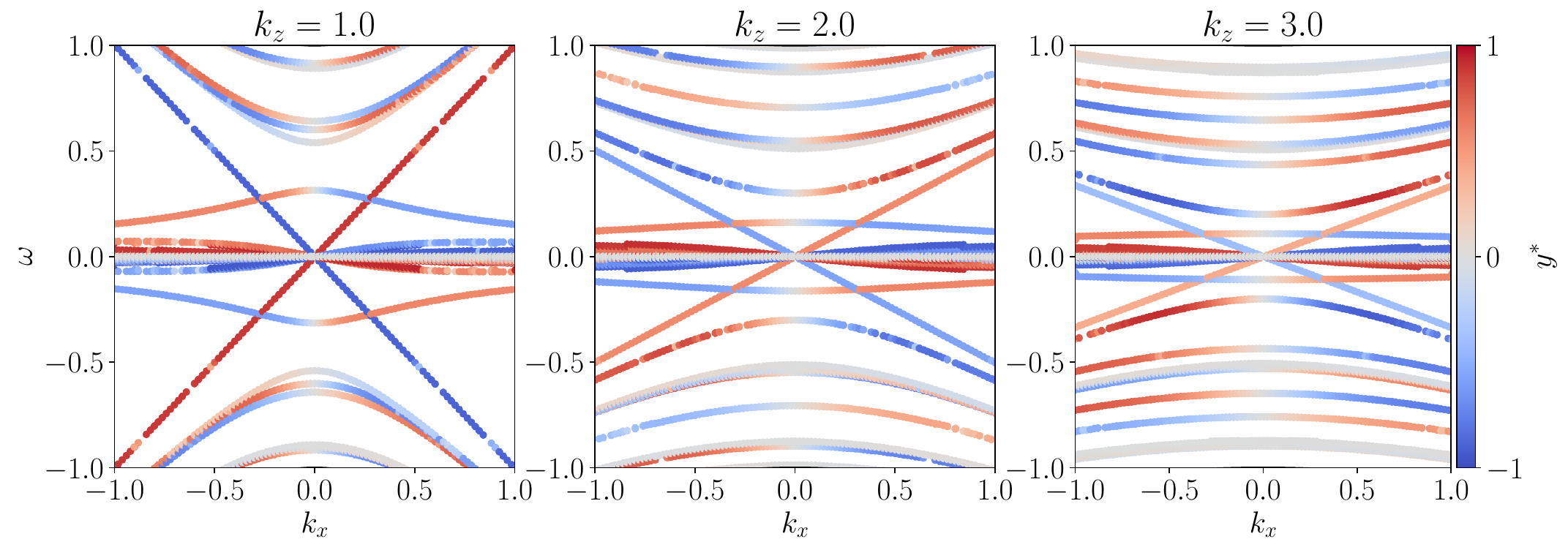} 
    \caption{The spectral flow of Kelvin and Yanai waves between the band gaps exhibited by the linearized primitive equations with a linear vertical shear $U_0 = 0.01$ obtained from \texttt{Dedalus} with $N_y = 24, N_z = 24$, $L_y = 10\pi, L_z = 2\pi$ using Fourier basis. The vertical wavenumber is (a) $k_z = 1$; (b) $k_z = 2$; and (c) $k_z = 3$. The solid black lines are the dispersion relation for the f-plane approximation with $f = 1$,  Eq.~\eqref{eqn:primitive_dispersion}. As in Fig.~\ref{fig:dedalus}, the color indicates proximity to the two equators. The missing scattered points are due to the difficulty in separating out the modes that correspond to different vertical wavenumbers $k_z$. }
    \label{fig:dedalus_pe}
\end{figure*}

With a vertical shear flow, the primitive equations are modified to be the following: 
\begin{eqnarray}
    \pdv{u}{t}  = -U(z) \pdv{u}{x} - f(y) v - \pdv{\eta}{x}, \nonumber \\
    \pdv{v}{t} = -f(y) u - U(z) \pdv{v}{x} - \pdv{\eta}{y}, \nonumber \\
    \pdv{t} \pdv{\eta}{z} = -w - U(z) \pdv{\eta}{x}{z}. \label{eqn:vertical_shear}
\end{eqnarray}
In this work, we consider a linear vertical shear flow, namely, $U(z) = U_0 z$. We numerically simulated the spectra in the $(y,z)$ space using \texttt{Dedalus} and is shown in Fig.~\ref{fig:dedalus_pe} with $U_0 = 0.05$. Similar to Fig.~\ref{fig:dedalus}, we use a sinusoidal Coriolis parameter $f(y) = \sin(2\pi y/L_y)$ with $L_y = 10\pi$. Both Yanai and Kelvin waves are present at different values of vertical wavenumber $k_z$.

\section{Shear induced instability} \label{sec:instability}
\begin{figure}[ht!]
\centering
     \includegraphics[width=0.5\linewidth]{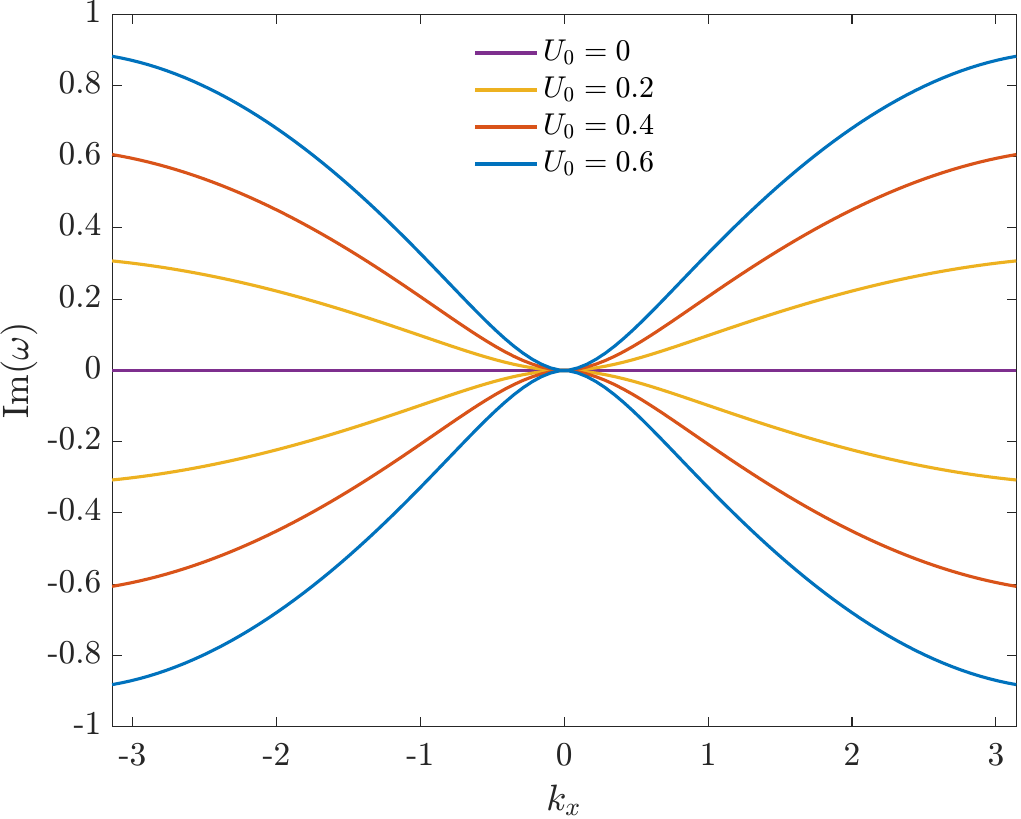} 
    \caption{Imaginary part of the frequency of the lowest-frequency planetary waves obtained from full diagonalization of the $69\times69$ linear wave operator. }
    \label{fig:imag}
\end{figure}
To investigate the stability of the waves in the presence of horizontal shear we follow Ref.~\cite{ripa1983general}.  Introducing the background potential vorticity $Q(y) = \frac{f(y) - \partial_y U(y)}{H(y)}$, perturbations are bounded if there exists some constant $\alpha \in \mathbb{R}$ such that the following two conditions hold for all $y \in \left[ -\frac{L_y}{2}, \frac{L_y}{2} \right]$:
(i) $[ \alpha - U(y)]~ \partial_y Q(y) \geq 0$ and $[\alpha - U(y)]^2 \leq H(y)$. 
For the sine horizontal shear flow condition (ii) can be satisfied, but condition (i) requires that the function
$g(U_0, y) = U_0 \sin (2\pi y)$ 
to be greater or equal to zero over the entire domain, but this condition is violated for any  $U_0 \neq 0$.  The analysis is similar with a cosine shear. Thus the bulk modes are always unstable in the presence of horizontal shear. 
We numerically confirm the instability of the bulk modes by presenting the imaginary part of the frequency spectrum in Fig.~\ref{fig:imag}. When $U_0\neq 0$, the spectrum has a non-zero imaginary part that grows linearly in $U_0$ for small shear.  The instability is most prominent in the planetary Rossby modes. 

In the presence of the linear vertical shear with rigid-lid boundaries, since the derivative of $U(z)$ has the same sign at the upper and lower boundaries, Eady instabilities are present at low wavenumbers ~\cite{Vallis2017}. 
We numerically verified the presence of Eady instabilities by simulating the primitive equations and observed that the spectra are unstable at low wavenumbers. 

Despite the presence of instabilities with both horizontal and shear flows, the gauge-invariant phase is a robust method of quantifying the topological nature of the system.

\section{Numerical Calculation of Bulk Winding Numbers}~\label{sec:winding} 

For Hermitian systems, bulk-interface correspondence~\cite{hatsugai1993chern,hasan2010topological} establishes a relationship between the topological invariant, the Chern number of the bulk, and the number of edge modes. 
It states that that the difference in the number of counterpropagating edge modes equals the difference in the Chern number in two bulk regions that are connected at a boundary: $\Delta C = n_L - n_R$, where $n_L$ and $n_R$ are the number of left-moving and right-moving modes. 
The Chern number can be calculated analytically for the rotating shallow water equations. Each of the 3 bands may be parametrized on the unit $(k_x, k_y, f)$ sphere.  The Chern number may then be found by integrating the Berry curvature over the surface of the sphere with a fixed radius $\sqrt{k^2 +f^2}$~\cite{delplace2017}.

In the presence of shear, however, the linear wave operator is no longer Hermitian, and a rigorous bulk-interface correspondence principle is not in hand. %(see however \cite{Delplace2021}).  
We may still investigate the topological properties of the bulk wavefunctions and compare with the boundary mode spectrum to test whether or not bulk-interface correspondence continues to operate. However, the presence of shear breaks translational invariance in the $y$-direction and the integral of the Berry curvature becomes difficult to evaluate. As an alternative, we instead look for singularities in the phase of the wavefunctions which appear as vortices in wavevector space~\cite{wang2016vortex}.
In the context of polarization physics, it has been shown that the winding of the polarization azimuth, or the wavefunction phase, equals the enclosed Chern number ~\cite{fosel2017,Bouteyre.2022}. 
We set the Coriolis parameter such that it is in the bulk (namely, it does not change sign), and examine the phase of the wavefunctions in $(k_x, k_y)$ to check whether there is a vortex or antivortex in the phase. 

\subsection{Spatially varying Coriolis parameter}\label{sec:coriolis}
\begin{figure*}[ht!]
\centering
    \includegraphics[width=1.0\linewidth]{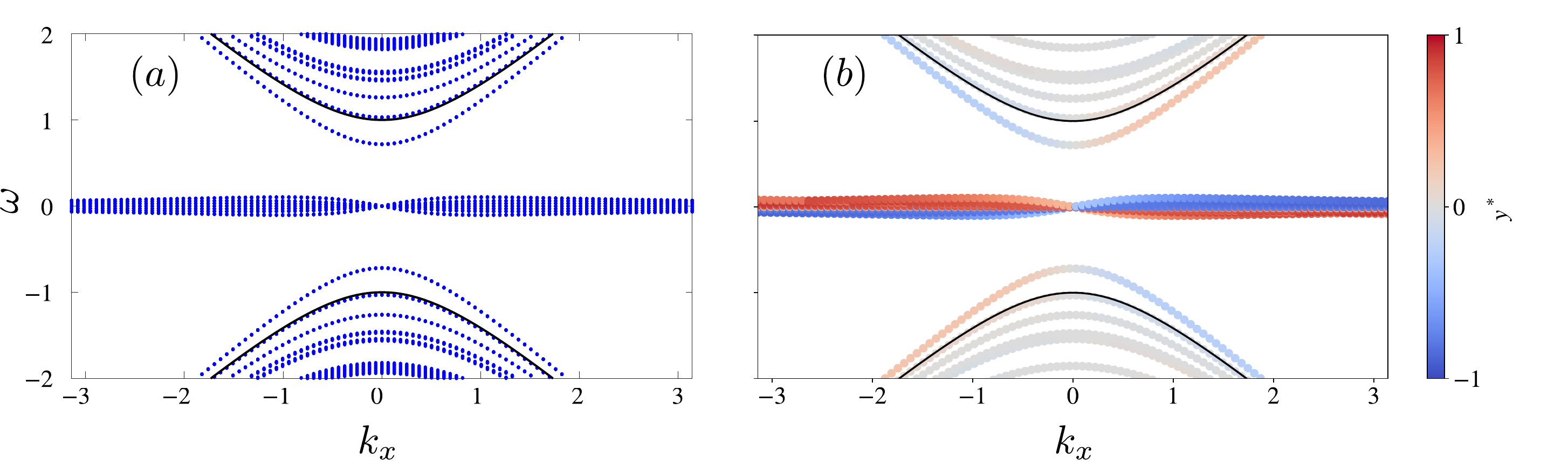}
    \caption{Numerical calculation of the bulk eigenfrequencies for the spatially varying Coriolis parameter. (a) Diagonalization of the $69 \times 69$ wavevector space linear wave operator.  (b) \texttt{Dedalus} with $N_y = 23$ for $\Delta f = 0.5$, $f_0 = 1$ and $L = 4\pi$. Black dotted lines in (a) and solid lines in (b) represent the frequency of Poincar\'e modes in the f-plane approximation with $f_0=1$: $\omega = \pm \sqrt{k_x^2 + f_0^2}$. Colors in (b) indicate proximity to the two oppositely oriented equators.  }
    \label{fig:fy_sin_spectrum}
\end{figure*}

\citet{delplace2017} uses an f-plane approximation to analytically calculate the Chern number to show the nontrivial topology of the equatorial waves. 
However, realistically, Coriolis parameter is a function of the latitude and translational invariance is always broken in the bulk. 
Here, we first verify that translational invariance in the bulk is not required. 
To do this we preserve Hermiticity by considering a spatially varying Coriolis parameter in absence of the shear flow and find the winding number of the Poincar\'e modes. 
We choose
\begin{eqnarray}
f(y) = f_0 + \Delta f \sin(\frac{2\pi y}{L_y}), \label{eqn:coriolis}
\end{eqnarray} 
so that we may adapt the formalism introduced in Eq.~\eqref{eqn:full} to write the linear wave operator in wavevector space with transition blocks: 
\begin{eqnarray} 
T_1(\Delta f) = \frac{\Delta f}{2} \begin{pmatrix}
0 & 1  & 0 \\ 
-1 & 0 & 0 \\ 
0 & 0 & 0
\end{pmatrix}, \quad
T_2( \Delta f) =T_1 (k_x, k_y, \Delta f) ^T.
\end{eqnarray}
By diagonalizing the linear wave operator, we can obtain the spectrum of shallow water equations with the $y$-dependent $f(y)$.  We choose $\Delta f$ and $f_0$ such that $f(y)$ does not change 
sign anywhere; thus we remain in the bulk and no edge modes should arise. 
Figure~\ref{fig:fy_sin_spectrum} shows the bulk spectrum with $\Delta f = 0.5$ and $f_0 = 1$. Frequencies obtained by diagonalization (Fig.~\ref{fig:fy_sin_spectrum}(a)) in wavevector space agree with those obtained with \texttt{Dedalus} (Fig.~\ref{fig:fy_sin_spectrum}(b)) and confirm that there are no Kelvin or Yanai waves.  

\subsection{Gauge invariant phase}
\begin{figure}[ht!] 
\centering 
	\includegraphics[width=0.8\linewidth]{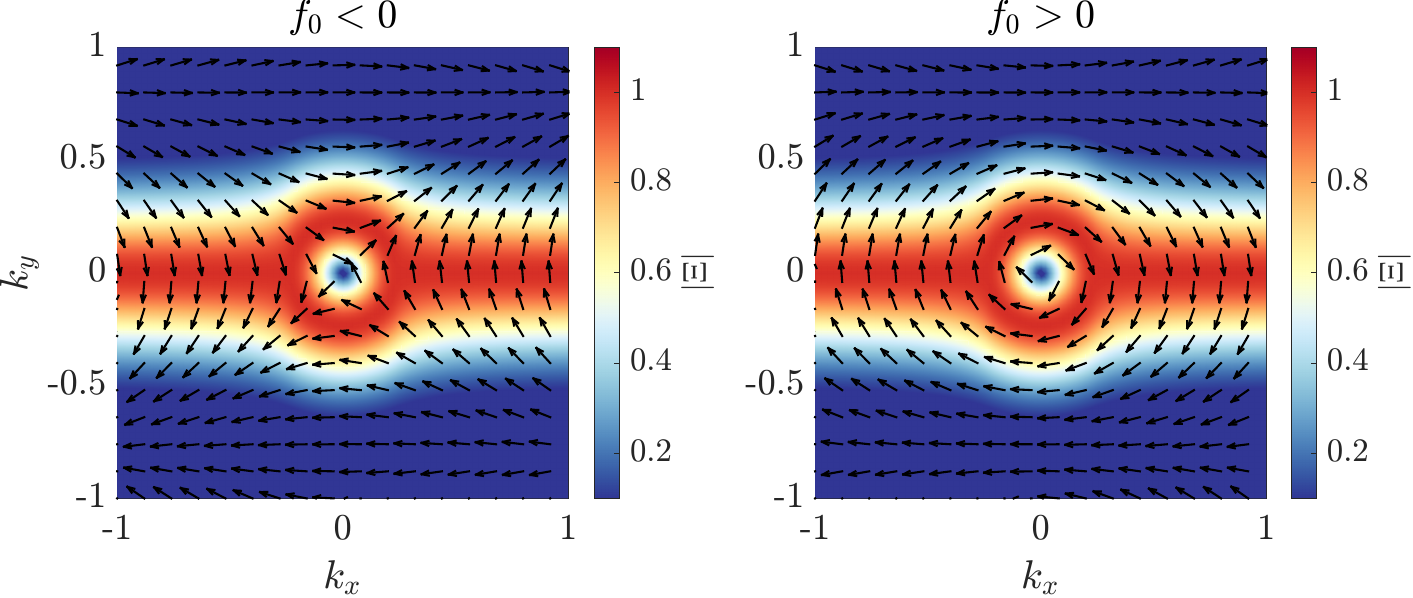}
	\caption{Arrows representing argument of $\Xi_\pm(\vec{k}) = v_\pm(\vec{k}) \eta_\pm(-\vec{k})$ of the lowest positive frequency Poincar\'e modes as indicated by the direction of the arrows, in the absence of shear but with the sinusoidal Coriolis parameter Eq.~ \eqref{eqn:coriolis} with $f_0 = -1$ (left) and $f_0=1$ (right), $\Delta f = 0.5$, $\Lambda = 10\pi$, and $L_y \rightarrow \infty$. Colors represent normalized magnitude $|\Xi|$ in arbitrary units.}
	\label{fig:fy_phase}
\end{figure}

\begin{figure}[ht!] 
\centering 
	\includegraphics[width=0.8\linewidth]{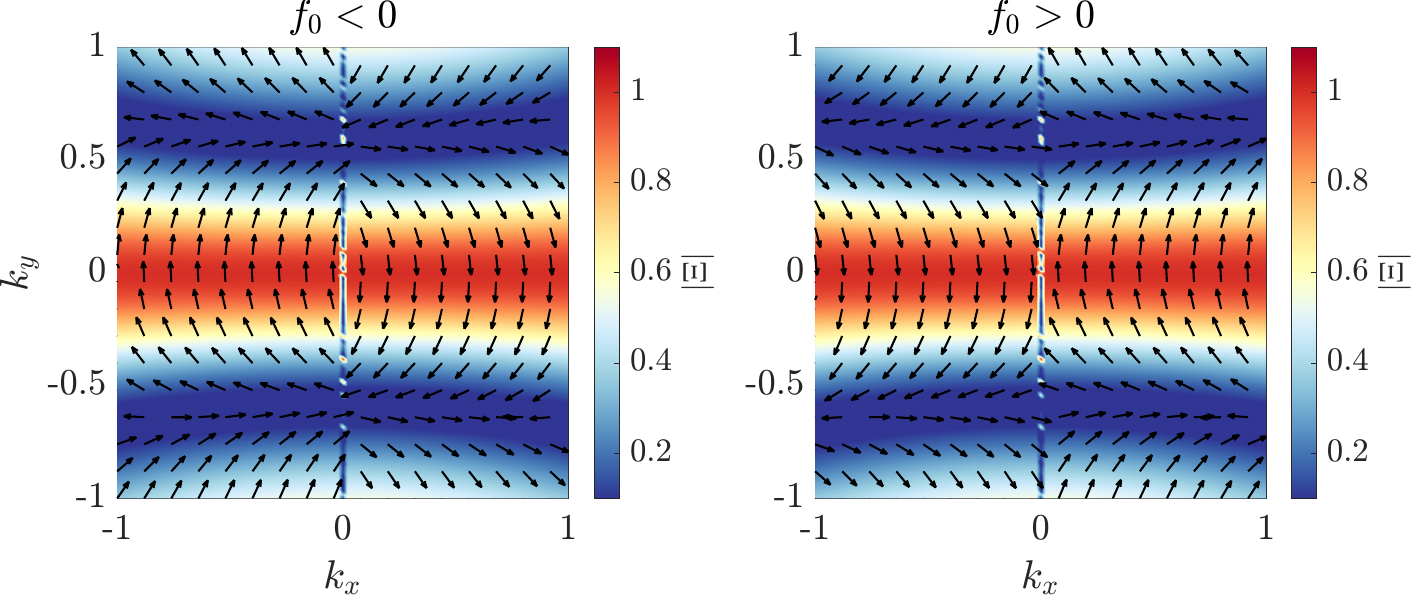}
	\caption{Arrows representing argument of $\Xi_\pm(\vec{k}) = v_\pm(\vec{k}) \eta_\pm(-\vec{k})$ of a Rossby mode as indicated by the direction of the arrows, in the absence of shear but with the sinusoidal Coriolis parameter Eq.~ \eqref{eqn:coriolis} with $f_0 = -1$ (left) and $f_0=1$ (right), $\Delta f = 0.5$, $\Lambda = 10\pi$, and $L_y \rightarrow \infty$. Colors represent normalized magnitude $|\Xi|$ in arbitrary units. Note that the precise pattern changes depending on which Rossby mode is chosen, but all cases are topologically trivial. } 
	\label{fig:fy_phase_rossby}
\end{figure}

We proceed to calculate the topological index of the bands by searching for singularities in the phase of the frequency eigenfunctions in wavevector space. The eigenfunctions have gauge freedom, as the phase of the three components can be rotated together by an arbitrary amount $\phi(\vec{k})$ at each point in wavevector space: 
\begin{eqnarray}
\Psi_{\pm, 0}(\vec{k}) \rightarrow e^{i \phi(\vec{k})} \Psi_{\pm, 0}(\vec{k}).
\end{eqnarray}
As mentioned previously in Section \ref{topology} we remove the gauge redundancy by multiplying the $v$-component of the Poincar\'e modes by the complex conjugate of the $\eta$-component, $\eta^*(\vec{k}) = \eta(-\vec{k})$:
\begin{eqnarray}
\Xi_\pm(\vec{k}) \equiv v_\pm(\vec{k})~ \eta_\pm(-\vec{k})
\end{eqnarray}
leaving only the internal phase difference between the two amplitudes.  
Figure~\ref{fig:fy_phase} depicts the argument of $\Xi_\pm(\vec{k})$, $\tan^{-1} (\mathrm{Re}(\Xi) /  \mathrm{Im} (\Xi))$, of the positive Poincar\'e modes as a function of $k_x$ and $k_y$ for the spatially varying Coriolis parameter of Eq.~\eqref{eqn:coriolis} where the eigenmodes are obtained by diagonalizing the $69 \times 69$ linear operator.  The positive Poincar\'e bands exhibit respectively a vortex and an anti-vortex centered at the origin in wavevector space where the phase cannot be uniquely defined for positive and negative $f_0$ respectively. The difference in the winding number between the two bands equals $2$. The difference in the winding number for either Poincar\'e band changes by $2$ going between the two hemispheres. The planetary waves have no vortex as expected (Fig.~\ref{fig:fy_phase_rossby}).

By virtue of the single-valuedness of $\Xi_\pm(\vec{k})$, the winding number must be integer-valued and thus topological in character.  Unlike the calculation of the Chern number which is found by integrating the Berry curvature over wavevector space, no integrals are required for the calculation of the winding number, and the non-compact nature of wavevector space for continuous fluids does not cloud its interpretation. 

The Chern number equals the negative of the total winding within a closed domain so $\Delta C = \nu_- - \nu_+$, where $\nu_\pm$ is the winding number of the positive/negative frequency Poincar\'e mode and a vortex/anti-vortex corresponds to a winding number $\pm 1$ ~\cite{fosel2017}. Thus $\Delta C_+ = -2$ for $f_0 > 0$ and  $\Delta C_- = 2$ for $f_0 < 0$, in agreement with the Chern numbers found for the f-plane~\cite{delplace2017}.  By bulk-interface correspondence~\cite{hatsugai1993chern,hasan2010topological}, the difference in the number of prograde and retrograde moving edge modes at the equatorial interface where $f$ changes sign equals the change in the Chern number $\{\Delta C_+, \Delta C_0, \Delta C_- \}$, consistent with $2$ modes of topological origin localized near each equator. The localized Yanai and Kelvin waves in Fig.~\ref{fig:dedalus}(a) thus have their origin in topology, just as they do for the shallow water equations using an f-plane approximation. \cite{delplace2017}. 

\subsection{Sinusoidal horizontal shear} 

\begin{figure}[ht!] 
\centering 
	\includegraphics[width=0.8\linewidth]{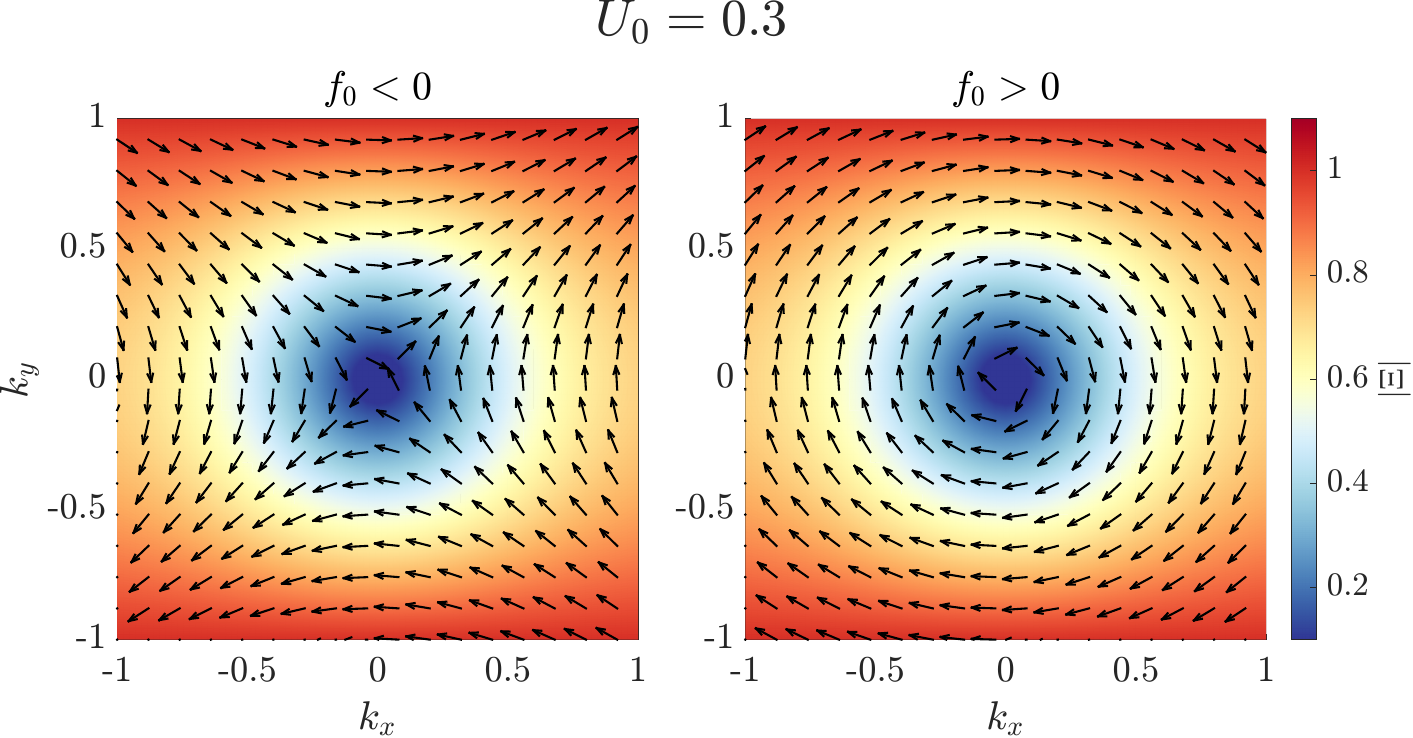}
	\caption{Arrows representing argument of $\Xi_\pm(\vec{k}) = v_\pm(\vec{k}) \eta_\pm(-\vec{k})$ of the lowest positive frequency Poincar\'e modes as indicated by the direction of the arrows for the case of sinusoidal horizontal shear $U_0 = 0.3$ within the f-plane approximation for $f_0 =-1 $ (left) and $f_0=1$ (right) with $L_y \rightarrow \infty$. The length of the arrows is rescaled to be equal. Colors represent normalized magnitude $|\Xi|$ in arbitrary units.}
	\label{fig:wf}
\end{figure}

Next we find the winding number of the Poincar\'e modes in the shallow water equations subjected to the sinusoidal horizontal shear. Figure~\ref{fig:wf} shows the phase of $\Xi_\pm({\bf k})$ for $U_0 = 0.3$ and constant Coriolis parameter $f_0=\pm 1$ showing qualitatively similiar vortices as those in Fig.~\ref{fig:fy_phase}. Again the positive frequency Poincar\'e modes exhibit a vortex for $f_0 > 0$ and an anti-vortex for $f_0 < 1$ at the origin in wavevector space (the phase singularity is absent for the planetary modes). The change in the winding number of $2$ is consistent with the number of edge modes seen in the spectrum (Fig.~\ref{fig:dedalus}(b) and (c)).  
This result suggests that the localized Kelvin and Yanai modes that traverse the gap between Rossby modes and the bulk Poincar\'e modes have a topological origin like the equatorial modes in the absence of shear.  This is the main result of the paper, and we note that 
the result also holds in perturbation theory with the $9\times 9$ linear wave operator, as the perturbative corrections to the wavefunction do not alter the winding number.  %We have also checked that the winding number remains unchanged in the presence of odd viscosity so long as the odd viscosity is chosen to break time-reversal symmetry in the same way as the Coriolis parameter.  
The appearance of Kelvin and Yanai waves along the equators shown in Section \ref{TwoEquators} is thus consistent with the persistence of the bulk-interface correspondence in the presence of shear.

% Fig.~\ref{fig:primitive v} presents the phase of the gauge-invariant quantity $\Xi_\pm({\bf k})$ for the linearized primitive equations with and without forcing from the sinusoidal horizontal shear. 
Finally, we study the phase of the gauge-invariant quantity $\Xi_\pm({\bf k})$ for the linearized primitive equations with and without forcing from sinusoidal horizontal shear. The phase singularity of $\Xi_\pm({\bf k})$ for primitive equations (not shown) is similar to that depicted in Figs.~\ref{fig:fy_phase} and \ref{fig:wf}. 
Without shear, the positive and negative frequency Poincar\'e modes have opposing winding numbers, and the winding number also changes polarity when $f$ changes sign in agreement with the analytic calculation of the Chern number. The vortex of the bulk Poincar\'e modes continues to be robust in the presence of shear, despite the combined effects of broken translational invariance, non-Hermiticity, and instability.  We have verified that the dispersion relation of the shear-forced primitive equations on the planet with two equators continues to exhibit spectral flow of the Kelvin and Yanai waves across the band gaps.  

\subsection{Linear vertical shear}
\begin{figure}[ht!] 
\centering 
	\includegraphics[width=0.6\linewidth]{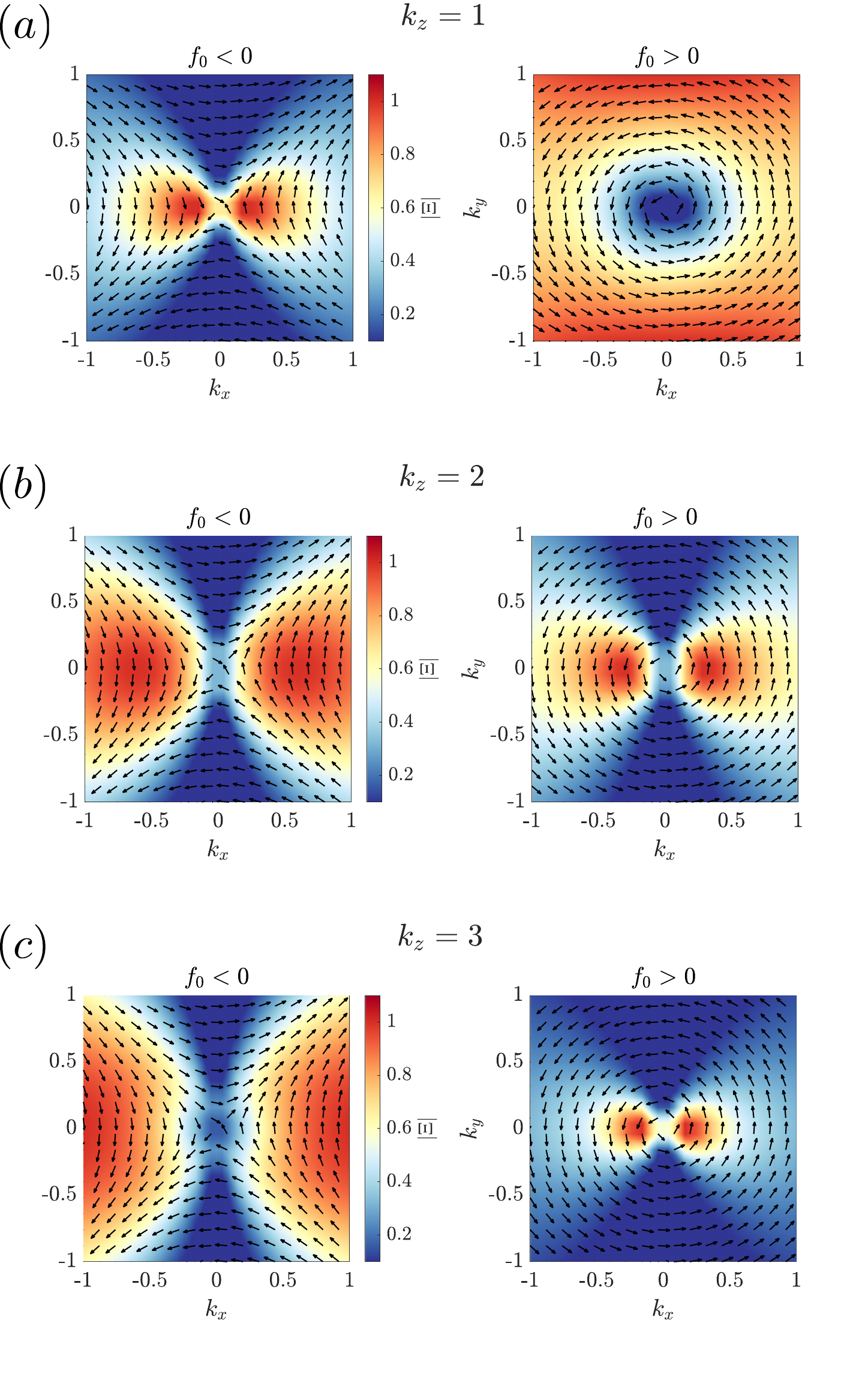}
	\caption{Arrows representing argument of $\Xi_\pm(\vec{k}) = v_\pm(\vec{k}) \eta_\pm(-\vec{k})$ of a positive Poincar\'e mode as indicated by the direction of the arrows for the case of a linear vertical shear flow with $U_0=0.05$ within the f-plane approximation for $f_0=-1$ (left) and $f=1$ (right). (a) $k_z = 1$ (b) $k_z = 2$ (c) $k_z=3$. Colors represent normalized $|\Xi|$ in arbitrary units. Obtained using \texttt{Dedalus} with $N_z=20$ and $L_z=2\pi$ using a Fourier basis. }
	\label{fig:phase_vshear}
\end{figure}

We proceed to calculate the winding number of the Poincar\'e modes in the primitive equators subject to the linear vertical shear flow. 
Primitive equations with a vertical shear flow $U(z)$ are given as Eq.~\eqref{eqn:vertical_shear}, which we simulate using  \texttt{Dedalus}. 
Figure~\ref{fig:phase_vshear} shows that in the presence of linear vertical shear flow, the bulk Poincar\'e mode exhibits phase singularity, and the winding number depends on the sign of the Coriolis parameters for all vertical wavenumbers $k_z$'s. 
This suggests that the Yanai and Kelvin waves in Fig.~\ref{fig:dedalus_pe} are topologically nontrivial. 
Note that while Fig.~\ref{fig:phase_vshear} is obtained with a Fourier basis and thus the effect of the rigid-lid boundary is removed, we verified that the winding numbers are similar to Fig.~\ref{fig:phase_vshear} with a no-slip boundary condition using a Chebyshev basis for each coefficient. 
Therefore, the topological nature of the boundary waves is robust against the presence of the Eady instability.

\section{Discussion and Conclusion}~\label{sec:conclusion}
We investigated the topological properties of rotating shallow water equations and stratified primitive equations in the presence of shear flow that breaks translational invariance in the meridional direction and Hermiticity and introduces instabilities. The winding number of the phase of $\Xi_\pm({\bf k})$ serves as a convenient probe of topological properties of the wavefunctions. 
This alternative to calculating the Chern number remains computationally tractable in the absence of translational invariance and Hermiticity. It may find application to experimental and observational data as well as to idealized theoretical models such as those studied here, as it can be obtained from the (usually neglected) phase information of the cross-periodogram between different fields such as the zonal velocity and geopotential height.  
To verify that the method yields sensible results, we studied the bulk modes in the presence of a spatially varying Coriolis parameter that does not change sign and demonstrated consistency with the standard calculation of the Chern number on the f-plane~\cite{delplace2017}. 
An alternative and equivalent way of quantifying the topological invariant is through the spectral index by counting the number of upward-going eigenvectors for increasing momentum, which is useful when we have access to the wavefunction ~\cite{Faure.2019,venaille2023ray}.

Our main result is that the winding number for both the shallow water equations and primitive equations remains unchanged in the presence of forcing by background shear flow.  The difference in the winding number of the Poincar\'e bands on opposite sides of the equator matches with the number of unidirectional waves localized at the equator, consistent with a topological origin for these forced Kelvin and Yanai waves. For the stratified primitive equations, there are topologically protected modes at each allowed vertical wavenumber in analogy to the physics of weak three-dimensional topological insulators.  

We note that we do not rigorously prove the bulk-interface correspondence for the shear flows, nor topological protection. However, we show that the bulk spectrum in f-plane approximation evolves smoothly with increasing $U_0$ and the phase singularities persist in both the numerically found eigenmodes and in low-order perturbation theory, at least if $U_0$ is not too large.  It may be possible to generalize the approach taken in Ref. \cite{Delplace2021} for frictionally damped shallow water waves to the problem of background shear.  That system, and the problems investigated here, are invariant under the combined operation of parity and time-reversal (PT).  We leave this, and an investigation of the maximum shear that will support equatorial waves of topological origin, for future work.

\acknowledgments{We thank Dan Borgnia, Deven Carmichael, Dung Nguyen, Steve Tobias, and Antoine Venaille for helpful discussions. Z.Z. is supported by the STC Center for Integrated Quantum Materials, NSF Grant No. DMR-1922172, ARO MURI Grant No. W911NF14-0247, NSF DMREF Grant No. 1922165, and a Stanford Science fellowship. J.B.M. is supported in part by a grant from the Simons Foundation (Grant No. 662962, GF) and by U.S. National Science Foundation Grants No. OIA-1921199 and No. OMA-1936221.}

\begin{appendices}
\section{Linearized shallow water equations in the presence of horizontal shear}
\label{sec:derivation}

We begin with the nonlinear shallow-water equations in the presence of rotation:
\begin{eqnarray}
	&& \pdv{\vec{u}_\mathrm{tot}}{t} + ( \vec{u}_\mathrm{tot} \cdot \grad) \vec{u}_\mathrm{tot} = -g \grad h - \vec{f}  \times \vec{u}_\mathrm{tot},  \nonumber \\ 
	&& \pdv{h}{t} + \grad \cdot (h \vec{u}_\mathrm{tot} ) = 0, \label{eqn:continuuity}
	\end{eqnarray} 
where $\vec{u}_\mathrm{tot} = \vec{u} + \vec{U}$, $\vec{u} = ({u}, {v})$, $\vec{U} = (U(y), 0)$ is the shear flow along the zonal direction, $\vec{f} = f(y) \hat z$ is the Coriolis parameter, and $h = \eta + H(y)$. 
To the linear order, Eq.~\eqref{eqn:continuuity} can be written as follows, 
\begin{widetext}
    \begin{eqnarray}
        && \pdv{u}{t} + U(y) \pdv{u}{x} + v\pdv{U(y)}{y} + g\pdv{\eta}{x}- f(y) v=0, \nonumber \\
        && \pdv{v}{t} + U(y) \pdv{v}{x} + g\pdv{\eta}{y} + f(y)  u = 0, \nonumber \\ 
        && \pdv{\eta}{t} + H(y) \left (\pdv{u}{x} + \pdv{v}{y} \right)  + v  \pdv{H(y)}{y} + U(y) \pdv{ \eta }{x}  = 0.
            \label{eqn:linear}                                                             
    \end{eqnarray}
\end{widetext}
A deformation length scale $L_d$ and gravity wave speed $c$ are defined to be:
\begin{eqnarray}
L_d \equiv \frac{c}{2\Omega}, \quad c \equiv \sqrt{g H}, 
\end{eqnarray} 
where $H$ is the zonally averaged depth without shear ($H(y) = H + h(y)$). 
Introducing the dimensionless quantities
$\tilde{t} = 2\Omega t$,  $\tilde{\eta} = \frac{\eta}{H}, \tilde{H}(y) = 1+\frac{h(y)}{H}$, $\vec{\tilde u} = \frac{u}{c}$,  $\vec{U} = \frac{U}{c}$, $\tilde{f} (y) = \frac{f(y)}{2\Omega}$, and $\vec{\tilde x} = \frac{\vec{x}}{L_d}$, the linearized equations of motion (Eq.~\eqref{eqn:linear}) around the basic state ($\vec{u} = 0, h = H$) can then be written as follows: 
\begin{eqnarray}
&& \partial_{\tilde t} \tilde u + \tilde U(y) \partial_{\tilde x} \tilde u + \tilde v \partial_{\tilde y} \tilde U(y) + \partial_{\tilde x} \tilde{\eta} -\tilde f(y) \tilde v  = 0, \nonumber \\
&& \partial_{\tilde t} \tilde v + \tilde U(y) \partial_{\tilde x} \tilde v + \partial_{\tilde y} \tilde \eta + \tilde f(y) \tilde u = 0,\nonumber \\
&& \partial_{\tilde{t}} \tilde{\eta} + \tilde{H}(y) (\partial_{\tilde{x}} \tilde{u} + \partial_{\tilde y} \tilde v) + \tilde v \partial_{\tilde y} \tilde H(y) + \tilde{U} (y) \partial_{\tilde x} \tilde{\eta} = 0. 
\label{eqn:swe_shear}
\end{eqnarray}
For convenience, we drop the tilde in the main text. 
%When $U(y) = 0$, 
%\begin{eqnarray}
%\partial_{t} u + \partial_{x} \eta - f v  = 0, \nonumber \\
%\partial_{t} v  + \partial_{y} \eta + f u = 0, \nonumber \\ 
%\partial_{t} {\eta} + \partial_{x} u + \partial_{y}  v = 0,
%\end{eqnarray} 
%which agrees with the shearless shallow water equations~\cite{delplace2017}.

%\section{Odd viscosity}\label{sec:odd}
%Odd viscosity regularizes the behavior of the linear wave operator at large %wavevectors\cite{Souslov:2019bl,Tauber:2020kf}. Odd viscosity $\nu_o$ may be easily included by %making the replacement $f \rightarrow f + \nu_o (k_x^2 + k_y^2)$ in $L_0$ (Eq. \ref{eqn:swe_h0}). % Figure~\ref{fig:odd_vis} compares the spectrum from full diagonalization with a small odd viscosity $\nu_o = 0.01$ to the spectrum from Dedalus. For a small $\nu_o$, the two spectra agree well. However, as $\nu_o$ increases, the two starts to diverge especially near the small $k_x$ regime. 
% \begin{figure*}[ht!] 
% \centering 
% 	\includegraphics[width=1.0\linewidth]{{dedalus_diagonalization_nu_o_0.01}.pdf}
% 	\caption{Comparison between the positive frequency Poincar\'e modes with a nonzero odd-viscosity $\nu_0 = 0.01$, from (a) full diagonalization with a $69 \times 69$ linear wave operator and (b) \texttt{Dedalus} with $N_y = 23.$ (c) The difference between $\delta \omega = \omega(U_0) - \omega(U_0=0)$ of the lowest positive Poincar\'e mode in (a) and (b).  }
% 	\label{fig:odd_vis}
% \end{figure*}

\section{The Shallow water linear wave operator in wavevector space}\label{sec:selection}

The matrix elements of the linear wave operator Eq.~\eqref{Lhat} in wavevector space may be written using Dirac braket notation as $\langle k_x^\prime, k_y^\prime | \hat{L} | k_x, k_y \rangle$.
Since the linear wave operator has no dependence on $x$ these matrix elements are non-zero only for $k_x^\prime = k_x$.
%the integral over $x$ simply returns a Dirac delta function $\delta(k_x^\prime-k_x)$. That is, 
%\begin{eqnarray}
%\int_{-\infty}^\infty \mathrm{d}x\, g(k_x) \mathrm{e}^{i (k_x^\prime - i k_x) x} = g(k_x) \delta(k_x^\prime - k_x),
%\end{eqnarray}
%for any function $g(x)$, which suggests that the states with different $k_x$s do not mix. 
Along the $y$-direction, we make use of the following relations,
\begin{eqnarray}
\frac{1}{L_y} \int_{-L_y/2}^{L_y/2} \mathrm{d}y~ \sin(\frac{2 \pi y}{L_y})~ \mathrm{e}^{i (k_y^\prime - k_y) y} 
= \frac{1}{2i} \left[\delta_{k_y^\prime, k_y - 2\pi/L_y} - \delta_{k_y^\prime, k_y + 2\pi/L_y}  \right],
\end{eqnarray} 
and 
\begin{eqnarray}
\frac{1}{L_y} \int_{-L_y/2}^{L_y/2} \mathrm{d}y~ \cos(\frac{2 \pi y}{L_y})~ \mathrm{e}^{i (k_y^\prime - k_y) y} 
= \frac{1}{2} \left[\delta_{k_y^\prime, k_y - 2\pi/L_y} + \delta_{k_y^\prime, k_y + 2\pi/L_y}  \right].
\end{eqnarray} 
In the absence of shear ($U_0 = 0$), the linear wave operator in $k$-space is a block-diagonal matrix, with the diagonal blocks being $L_0 (k_x, k_y, f)$ and with no off-diagonal blocks.  The $3\times3$ linear wave operators $L_0$ at wavevectors $(k_x, k_y)$ and $(k_x, k_y \pm 2\pi)$ are connected by the sinusoidal horizontal shear as a wave at wavevector $k_y$ mixes with modes $k_y^\prime = k_y \pm 2\pi$.  For a given $k_x$, we need to diagonalize the full matrix in the basis of $k_y, k_y \pm 2\pi, k_y \pm 4\pi, ...$ imposing a finite cutoff in $|k_y^\prime|$ to keep the dimension of the matrix finite.

\section{Comparison with Dedalus} \label{sec:dedalus}

\begin{figure}[ht!]
\centering
    \includegraphics[width=0.8\linewidth]{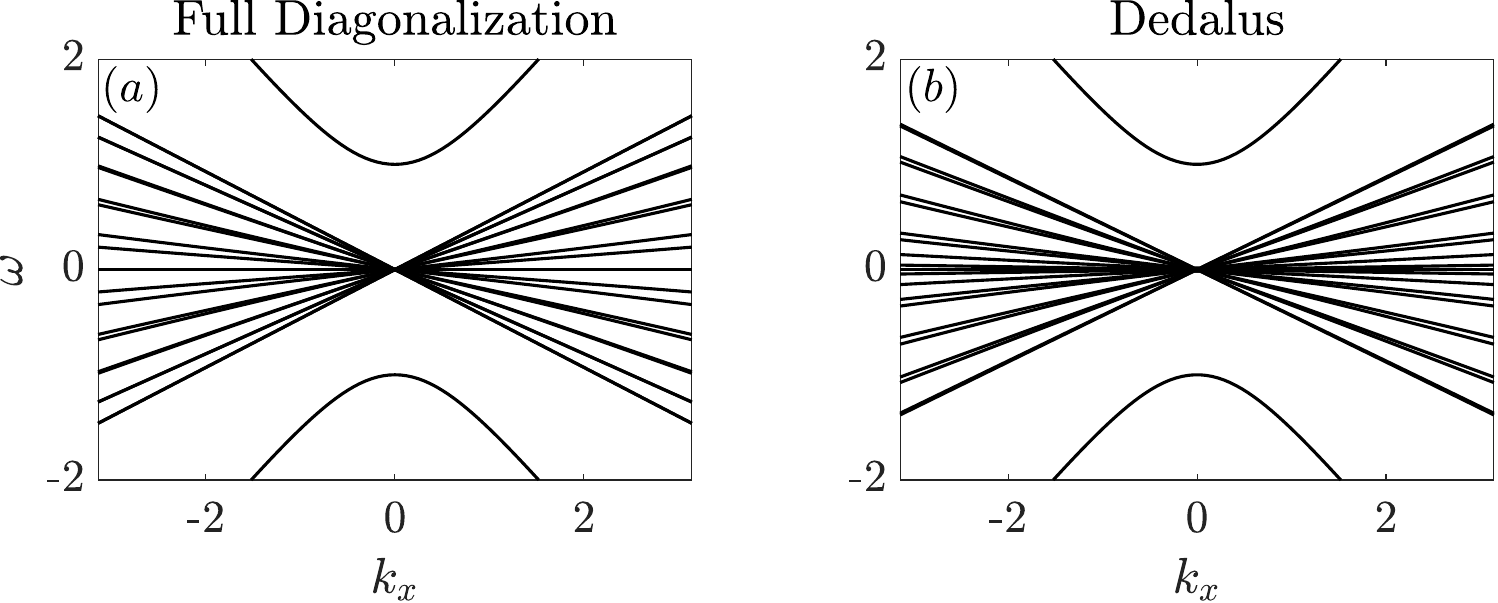} 
    \caption{Frequency spectra of the shallow water equations in the f-plane approximation with $f=1$ and subjected to sine shear $U_0=0.5$. The frequencies are obtained by (a) diagonalizing the $69\times69$ wavevector space linear wave operator and from (b) Dedalus with $N_y = 23$.}
    \label{fig:fplane}
\end{figure}

\begin{figure*}[ht!] 
\centering 
	\includegraphics[width=1.0\linewidth]{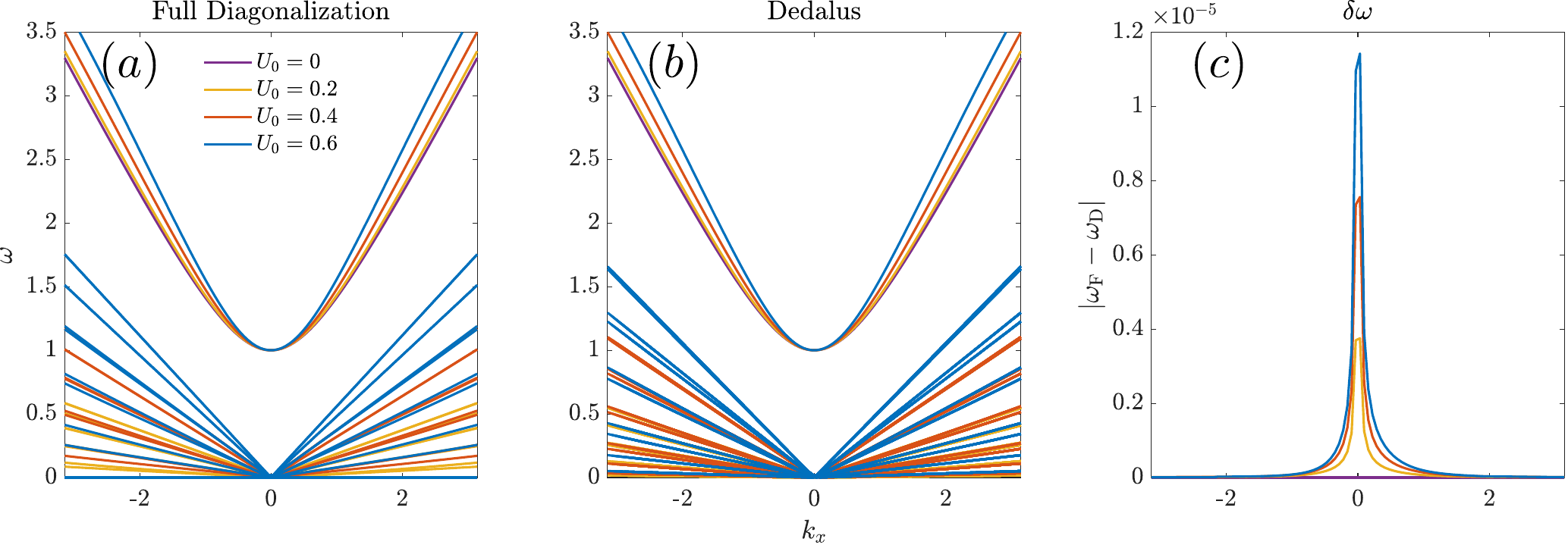}
	\caption{Comparison of the frequencies of the positive Poincar\'e and planetary modes. (a) Full diagonalization of the $69 \times 69$ linear wave operator. (b) \texttt{Dedalus} with $N_y = 23$. (c) The difference between frequencies of the lowest positive Poincar\'e mode obtained from full diagonalization, $\omega_\mathrm{F}$, and Dedalus, $\omega_\mathrm{D}$ in (a) and (b).}
	\label{fig:U0_sweeps}
\end{figure*}

We validate our diagonalization scheme by comparing with \texttt{Dedalus}~\cite{burns2019dedalus}. 
Figure~\ref{fig:fplane} compares the spectra from diagonalizing a $69 \times 69$ linear wave operator corresponding to the $23$ retained wavevectors in the $y$-direction with the spectrum obtained from \texttt{Dedalus}. To enable the comparison, the linear wave operator has been truncated to finite dimension in wavenumber space to match the total number of equations in \texttt{Dedalus}.  The full diagonalization captures both the spread of the geostrophic modes and the bulk Poincar\'e modes.
Note that the small difference in the geostrophic modes is due to the fact that the sample points along the $y$-direction in \texttt{Dedalus} is non-uniform whereas in the direct diagonalization, $k_y$'s are sampled uniformly.
Figure~\ref{fig:U0_sweeps} compares the positive frequency modes obtained from full diagonalization versus those found using \texttt{Dedalus}. The two methods show an excellent agreement. The frequency of the Poncar\'e modes increases with increasing shear and remain distinct beyond $U_0 = 0.6$.
We can apply the same procedure to obtain the transition matrices $T_1$ and $T_2$ for the cosine shear, and the spectra agrees with Figs.~\ref{fig:fplane} and \ref{fig:U0_sweeps}, as expected.

\section{Primitive equations} \label{sec:pe_w_shear}

Using the same non-dimensionalization as Appendix~\ref{sec:derivation}, the non-dimensional Boussinesq primitive equations are given as follows~\cite{Vallis2017}:
\begin{eqnarray}
&& R_0\frac{D{\textbf{\emph{u}}}}{D{t}} + \textbf{\emph{f}}(y)\times {\textbf{\emph{u}}} = -\grad {\phi}, \nonumber \\
&& R_0\frac{D \emph{b}}{D t}+\left(\frac{L_d}{L_y}\right)^2 N^2 w = 0 \nonumber \\
&& \partial_{z} \phi = {b},\nonumber \\
&& \partial_{x} {\emph{u}} + \partial_{{y}} \emph{v} + \partial_{z} \emph{w} = 0.
\label{eqn:primitive eq}
\end{eqnarray}
where $w$ is the vertical velocity, $\phi$ is the kinetic pressure, and $b$ is the buoyancy fluctuation about an average stratification, $N^2 = \partial b/\partial z$, and $L_d$ is the deformation radius, and $R_0$ is the Rossby number.
We consider the linearized equations
\begin{eqnarray}
&& R_0\frac{\partial {\textbf{\emph{u}}}}{\partial t} + \textbf{\emph{f}}(y) \times \textbf{\emph{u}}  = -\grad \phi, \nonumber \\
&& R_0\frac{\partial \emph{b}}{\partial t}+\left(\frac{L_d}{L_y}\right)^2 N^2 w = 0, \nonumber \\
&& \partial_{ z}  \phi = b,\nonumber \\
&& \partial_{x} \emph{u} + \partial_{y} \emph{v} + \partial_{z} \emph{w} = 0.
\label{eqn:linearized primitive eq}
\end{eqnarray}
Here, $R_0$ and $N L_d / L_y$ can be set to unity by appropriate re-scaling of the variables.
In the Fourier space, $-ik_z\phi = b$ and $ik_x u + ik_y v+ik_z w = 0$. Therefore, we can eliminate $\phi$ and $w$ by writing them in terms of $b$, $u$ and $v$. In the f-plane approximation, the dispersion relation for the Poincar\'e modes is $\omega^2 = f^2+(k_x^2+k_y^2)/k_z^2$.

Finally we consider the imposition of sinusoidal horizontal shear flow. We assume the system is periodic in the zonal and meridional direction and has rigid lids at $z=0$ and $z=L_z$, where $L_z$ is a constant. Let $\vec{u}_\mathrm{tot} = \vec{u} + \vec{U}$, $\vec{u} = ({u}, {v})$, $\vec{U} = (U(y), 0)$ and $\phi = \eta + H(y)$.
From geostrophic balance, $U(y)$ and $H(y)$ must satisfy Eq.~\eqref{eqn:balance}. Substituting $\vec{u}_\mathrm{tot}$ and $\phi$ into Eq.~\eqref{eqn:primitive eq} and discarding non-linear terms, we obtain:
\begin{eqnarray}
&& \frac{{\partial{u}}}{\partial{t}} = -U(y) \frac{\partial{u}}{\partial{x}} - v \frac{\partial{U(y)}}{\partial{y}} + \frac{f(y)}{R_0} v - \frac{1}{R_0} \frac{{\partial{\eta}}}{\partial{x}}, \nonumber \\
&& \frac{{\partial{v}}}{\partial{t}} =  -\frac{f(y)}{R_0} u - U(y) \frac{\partial{v}}{\partial{x}} - \frac{1}{R_0} \frac{{\partial{\eta}}}{\partial{y}},\nonumber \\
&& \frac{\partial}{\partial{t}} \frac{\partial{\eta}}{\partial{z}} = - \frac{1}{R_0} (\frac{L_d}{L_y})^2 N^2 w - U(y) \frac{\partial^2}{\partial{x} \partial{z}} \eta.
\label{eqn:primitive eq with shear}
\end{eqnarray}
Again $R_0$ and $N^2(L_d/L_y)^2$ may be set to unity. By doing so, Eq.~\eqref{eqn:primitive eq with shear} simplifies to
\begin{eqnarray}
&& \frac{{\partial{u}}}{\partial{t}} = -U(y) \frac{\partial{u}}{\partial{x}} - v \frac{\partial{U(y)}}{\partial{y}} + f(y) v -  \frac{{\partial{\eta}}}{\partial{x}}, \nonumber \\
&& \frac{{\partial{v}}}{\partial{t}} =  -f(y) u - U(y) \frac{\partial{v}}{\partial{x}} -  \frac{{\partial{\eta}}}{\partial{y}},\nonumber \\
&& \frac{\partial}{\partial{t}} \frac{\partial{\eta}}{\partial{z}} = - w - U(y) \frac{\partial^2}{\partial{x} \partial{z}} \eta.
\label{eqn:primitive eq with shear simplified}
\end{eqnarray}

\end{appendices}

\bibliography{shear_reference}	
\end{document}